\documentclass[12pt]{article}

\usepackage[dvips]{graphicx}

\newcommand{\lsim}{\raisebox{-4pt}{$\,\stackrel{\textstyle
                                                         <}{\sim}\,$}}

\newcommand{\nn}{\nonumber}
\newcommand{\be}{\begin{equation}}
\newcommand{\ee}{\end{equation}}
\newcommand{\ba}{\begin{eqnarray}}
\newcommand{\ea}{\end{eqnarray}}
\newcommand{\req}[1]{(\ref{#1})}
\def\={\,=\,}
\newcommand{\ci}[1]{\cite{#1}}

\def\mev{~{\rm MeV}}
\def\gev{~{\rm GeV}}

\def\als{\alpha_{\rm s}}
\def\eps{\epsilon}

\def\xbj{x_{\rm Bj}}

\newcommand{\tw}{\textwidth}
\def\vk{{\bf k}_{\perp}}

\def\vbs{{\bf b}}
\def\vb0{{\bf b}_0}

\def\xbj{x_{\rm Bj}}
\newcommand{\sla}{\hspace*{-0.015\tw}/}

\newcommand{\da}{{distribution amplitude}}
\newcommand{\wf}{wave function}

\def\={\,=\,}

\begin{document} 
\thispagestyle{empty}
\begin{flushright}
WU B 13-11 \\
October, 02 2013\\[20mm]
\end{flushright}

\begin{center}
{\Large\bf Transversity in exclusive vector-meson leptoproduction} \\
\vskip 10mm

S.V.\ Goloskokov
\footnote{Email:  goloskkv@theor.jinr.ru}
\\[1em]
{\small {\it Bogoliubov Laboratory of Theoretical Physics, Joint Institute
for Nuclear Research,\\ Dubna 141980, Moscow region, Russia}}\\
\vskip 5mm

P.\ Kroll \footnote{Email:  kroll@physik.uni-wuppertal.de}
\\[1em]
{\small {\it Fachbereich Physik, Universit\"at Wuppertal, D-42097 Wuppertal,
Germany}}\\
and\\
{\small {\it Institut f\"ur Theoretische Physik, Universit\"at
    Regensburg, \\D-93040 Regensburg, Germany}}\\
\vskip 5mm
\end{center}
\vskip 5mm 

\begin{abstract}
The role of transversity or helicity-flip generalized parton distributions (GPDs) in 
leptoproduction of vector mesons is investigated within the framework of the handbag approach.
The transversity GPDs in combination with twist-3 meson \wf s, occur in the amplitudes
for transitions from a transversely polarized virtual photon to a longitudinal polarized
vector meson. The importance of the transversity GPDs can be examined in some of the spin 
density matrix elements (SDMEs) and in transverse target spin asymmetries. Using suitable 
parametrizations of both helicity-flip and non-flip GPDs, which are essentially taken from 
our previous papers, we estimate these observables and compare the results with available data.

\end{abstract}

\section{Introduction}
\label{sec:intro}
While, in the framework of the handbag approach, the role of the helicity non-flip 
GPDs, $H, E, \widetilde{H}$ and $\widetilde{E}$, in deeply virtual Compton scattering 
and in exclusive meson leptoproduction have intensively been studied during the last 
fifteen years, the applications of the transversity or helicity-flip GPDs are rare. 
Only a few publications on this issue can be found in the literature, e.g.\ 
\ci{ji98}-\ci{GK6}. This is in sharp contrast to the situation of transversity in 
semi-inclusive reactions where a rich literature exists, see for instance the review 
articles \ci{barone,eic}. The reason for this fact is that, for the quark transversity 
GPDs, the emitted and reabsorbed partons have opposite helicities. Since the 
interactions of light quarks with gluons or photons conserve helicity, the initial 
parton helicity flip can only be compensated by higher-twist meson \wf s. Therefore, 
the contribution from the quark transversity GPDs are small in most cases and are 
difficult to separate from those of the helicity non-flip GPDs. For the gluon 
transversity GPDs the situation is different but it seems that their contributions 
are even smaller.

Leptoproduction of pseudoscalar mesons is an exception. On the one hand, the contributions
from $\widetilde H$ and $\widetilde E$  are rather small in this case. On the 
other hand, those from the transversity GPDs are comparably large since their contributions
are enhanced by the chiral condensate which appears in the \wf{}  for a (ground state)
pseudoscalar meson \ci{GK5}. This fact entails the dominance of the amplitudes for 
the transitions from a transversely polarized virtual photon to the pseudoscalar meson, 
$\gamma^*_T\to P$. The asymptotically leading amplitudes for the transitions from a 
longitudinally polarized photon, $\gamma^*_L\to P$, are much smaller according to the 
estimates made in \ci{GK5,GK6}. The only substantial contributions to these amplitudes 
are the meson-pole terms as, for instance, the pion pole in $\pi^+$ leptoproduction~\footnote{
The pion-pole contribution dominates the $\pi^+$ cross section at small momentum transfer
as is well-known. However, this result cannot be considered as a success of the handbag 
approach. A calculation of the $\pi^+$ cross section from LO Feynman graphs 
(see Fig.\ \ref{fig:handbag}) underestimates it markedly.}.
 
Here, in this work, we are going to investigate the role of the transversity GPDs in 
vector-meson leptoproduction. We will utilize the parametrizations of the helicity 
non-flip GPDs advocated for in \ci{GK3} as well as those of the valence-quark
transversity GPDs used in our study of leptoproduction of pseudoscalar mesons \ci{GK5,GK6}.
In addition we will allow for sea-quark contributions from these GPDs. As in \ci{GK5,GK6}
we will not perform detailed fits to experimental data. In so far the results we will
present below are to be understood as estimates. A more exact 
determination of the transversity GPDs is to be left for future investigations. 
Prerequisite to such an analysis are data on, say, the $\pi^0$ cross section at reasonably 
large photon virtuality, $Q^2$, and large c.m.s.\ energy, $W$. Such data may come from the 
COMPASS experiment or the upgraded Jefferson Lab.

The plan of the paper is the following: In the next section we will outline the handbag
approach, referring to our previous work \ci{GK5,GK6,GK3,GK1} and giving only details
for the treatment of the contributions from the transversity GPDs. In this section
we will also discuss the calculation of the subprocess amplitude for quark helicity flip
and present the parametrizations of the GPDs. In Sect.\ \ref{sec:results} we will
present our results for those observables of vector-meson leptoproduction which are
sensitive to the transversity GPDs. The paper is closed with a summary.  

\section{The handbag approach}      
\label{sec:handbag}

We consider the process $\gamma^*(q,\mu)\, p(p,\nu)\to V(q^\prime,\mu')\, p(p^\prime,\nu')$ 
in the generalized Bjorken-regime of large $Q^2$ and large $W$ but fixed Bjorken-$x$, 
$\xbj$. The symbols in the brackets denote the 
momenta and the helicities of the particles. The square of the momentum transfer, 
$\Delta=p^\prime-p$, is assumed to be much smaller than $Q^2$ ($t=\Delta^2$). 
We also restrict ourselves to small values of $\xbj$, i.e. to values of skewness,
\be
  \xi\=\frac{(p-p^\prime)^+}{(p+p^\prime)^+}\simeq \frac{\xbj}{2-\xbj}\,(1+m_V^2/Q^2)\,,
\ee
smaller than about 0.1 ($m_V$ denotes the mass of the vector meson $V$).
We stress that throughout the paper we neglect terms which are suppressed
as $\sqrt{-t}/Q$ or stronger. We will work in a photon-proton center-of-mass system 
where the proton momenta are defined as $p=\bar{p}-\Delta/2$ and $p^\prime=\bar{p}+\Delta/2$. 
The average proton momentum is $\bar{p}=(p+p^\prime)/2$ and we choose its three-momentum 
part to point along the 3-axis.

\begin{figure}
\centerline{\includegraphics[width=0.5\tw]{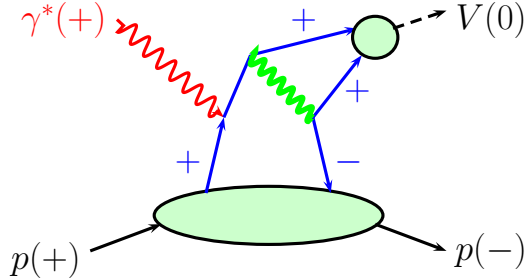}}
\caption{A typical graph for meson leptoproduction. The helicity labels
refer to the amplitude ${\cal M}_{0-,++}$ and to the subprocess $\gamma^*q\to (q\bar{q})q$.}
\label{fig:handbag}
\end{figure}

As described in detail in \ci{GK3,GK1} a helicity amplitude ${\cal M}_{\mu\nu',\mu\nu}$ 
is assumed to factorize in a hard subprocess amplitude $H_{\mu\lambda,\mu\lambda}$
(where $\lambda$ is the helicity of the internal partons, quarks or gluons)
and a soft proton matrix element, parametrized in terms of
GPDs, see Fig. \ref{fig:handbag}. Since the partons which are emitted 
and reabsorbed from the proton collinearly to its initial and final state momentum, 
have the same helicity in this subprocess amplitude the GPDs $H$ and $E$ appear in 
the convolution. There are, however, also small, nearly negligible contributions from 
$\widetilde{H}$ and $\widetilde{E}$ to the $\mu=\pm 1$ amplitudes. 

The subprocess amplitudes are calculated within the modified perturbative
approach \ci{li-sterman} in which quark transverse degrees of freedom in the
subprocess as well as Sudakov suppressions are taken into account. This
entails the necessity to use a light-cone wave function for the meson
instead of a distribution amplitude. In the limit of $Q^2, W \to\infty$
the subprocess amplitudes for transitions from a longitudinally polarized
photon to a likewise polarized vector meson, $\gamma^*_L\to V_L^{\phantom{*}}$, can be
shown to turn into the collinear result, i.e.\ the familiar asymptotic factorization
formula emerges for the amplitude ${\cal M}_{0\nu',0\nu}$. The factorization 
of ${\cal M}_{0\nu',0\nu}$ has rigorously been proven to hold in the limit of 
$Q^2, W \to\infty$\ci{radyushkin96,collins}. The infrared singularities known to occur 
in the subprocess amplitudes for transversely polarized photons and mesons 
$H^V_{\pm\lambda,\pm\lambda}$ in collinear approximation, are regularized by the quark 
transverse momentum, $\vk$, in the modified perturbative approach. (Note that 
explicit helicities are labeled by their signs or by zero.)  
The $\gamma^*_T\to V_T^{\phantom{*}}$ amplitudes are therefore suppressed by 
$\sqrt{\langle k^2_\perp\rangle}/Q$ with respect to those for 
$\gamma^*_L\to V_L^{\phantom{*}}$ transitions~\footnote{
For a different treatment of $\gamma^*_T\to V_T^{\phantom{*}}$ transitions see \ci{pire11}.}. 
For further details of the handbag approach we refer to \ci{GK3,GK1}.

The role of the transversity GPDs \ci{ji98,diehl01}
$H_T$, $\bar{E}_T=2\widetilde{H}_T+E_T$, $\ldots$ in exclusive leptoproduction 
of pseudoscalar mesons has been investigated in \ci{GK5,GK6}. Since for these GPDs 
the emitted and reabsorbed partons have opposite helicities they only contribute to
the amplitudes for transversely polarized photons to the order of accuracy we are working. 
As discussed in \ci{GK5,GK6} the contributions from the transversity GPDs seem to be 
dominant in most of the pseudoscalar channels. For instance, the transverse cross section 
for $\pi^0$ production is estimated in \ci{GK6} to be about 10 times larger than the 
longitudinal cross section which seems to be in agreement with experiment \ci{clas12,kim13}. 

Here, in this work we are going to explore the role of the
transversity GPDs in vector-meson leptoproduction. In full analogy to the
case of pseudoscalar mesons the quark transversity GPDs contribute to the
amplitudes ${\cal M}^V_{0\nu',\pm\nu}$ for $\gamma^*_T\to V^{\phantom{*}}_L$ transitions:
\ba
{\cal M}^V_{0+,++} &=& \phantom{-}\frac{e_0}2\,\frac{\sqrt{-t^\prime}}{2m} 
                      \sum_a e_a{\cal C}_V^a  \int dx \sum_\lambda \Big[
                 2\lambda H^V_{0\lambda,+-\lambda}\,\big(\bar{E}_T^a -\xi\widetilde{E}_T^a\big)\nn\\ 
         &+& H^V_{0\lambda,+-\lambda}\,\big(\widetilde{E}_T^a-\xi E_T^a\big)\Big]\,,\nn\\
{\cal M}^V_{0+,-+} &=& -\frac{e_0}2\,\frac{\sqrt{-t^\prime}}{2m} 
                      \sum_a e_a{\cal C}_V^a \int dx \sum_\lambda \Big[
            2\lambda H^V_{0\lambda,+-\lambda}\,\big(\bar{E}_T^a -\xi\widetilde{E}_T^a\big)\nn\\ 
      &-& H^V_{0\lambda,+-\lambda}\,\big(\widetilde{E}_T^a-\xi E_T^a\big)\Big]\,,\nn\\
{\cal M}_{0-,++} &=& e_0\sqrt{1-\xi^2}\,\sum_a e_a{\cal C}_V^a  \int dx \Big[ 
       H^V_{0-++}\big(H_T+\frac{\xi}{1-\xi^2}\big(\widetilde{E}_T^a-\xi E_T^a\big)\big)\nn\\
    &+&\frac{t^\prime}{2m^2}\,\sum_\lambda \lambda H^V_{0\lambda,+-\lambda}\,\widetilde{H}_T^a\Big]\,,\nn\\
{\cal M}^V_{0-,-+} &=& e_0\sqrt{1-\xi^2}\, \sum_a e_a{\cal C}_V^a \int dx \Big[ 
       H^V_{0--+}\big(H_T^a+\frac{\xi}{1-\xi^2}\big(\widetilde{E}_T^a-\xi E_T^a\big)\nn\\
    &-&\frac{t^\prime}{2m^2}\,\sum_\lambda \lambda H^V_{0\lambda,+-\lambda}\big)\,\widetilde{H}_T^a\Big]\,.
\label{eq:TL}
\ea
As independent amplitudes we choose those with $\nu=1/2$. The amplitudes with
$\nu=-1/2$ are related to the other ones by parity conservation~\footnote{
This relation holds analogously for the subprocess amplitudes.}: 
\be
{\cal M}^V_{-\mu'-\nu',-\mu-\nu}\= (-1)^{\mu-\nu-\mu'+\nu'}
                                    {\cal M}^V_{\mu'\nu',\mu\nu}\,.
\ee
Since we neglect contributions which are suppressed at least by $\sqrt{-t}/Q$, 
only helicity-non-flip subprocess amplitudes can appear in the convolutions
\req{eq:TL}. For quark helicity-flip the only subprocess amplitude of this type is 
$H^V_{0-,++} (=H^V_{0+,--})$ and, hence, only the $\gamma^*_T\to V_L^{\phantom{*}}$ transitions 
are fed by the transversity GPDs to the order of accuracy we are working. The expressions 
\req{eq:TL} can easily be derived with the help of the proton-quark matrix elements 
given in \ci{diehl01}. In \req{eq:TL} $m$ is the proton mass, $a$ denotes the quark 
flavor and $e_a$  the quark charges in units of the positron charge, $e_0$. For unflavored
mesons the non-zero flavor weight factors, ${\cal C}_V^a$, read
\be
{\cal C}^u_{\rho^0}\=-{\cal C}^d_{\rho^0}\={\cal C}^u_{\omega}\=
{\cal C}^d_{\omega}\=1/\sqrt{2}\,, \qquad {\cal C}^s_\phi\=1\,.
\label{eq:flavor}
\ee
For the flavored mesons, $\rho^+$ and $K^{*0}$, the $p\to n$ and $p\to \Sigma^+$ transition
GPDs appear. As a consequence of isospin symmetry or SU(3) flavor symmetry the transition 
GPDs can be related to the corresponding proton GPDs \ci{frankfurt99}~\footnote{
The different masses of the nucleon and the hyperon are taken into account as
in \ci{GK4}.}
\be
K^{\rho^+} \= K^u - K^d\,, \qquad K^{K^{*0}}\= -K^d+K^s\,,
\label{eq:non-diagonal}
\ee 
where K is some GPD. For these mesons there are no flavor weight factors and the 
charges have to be absorbed into the subprocess amplitudes.
Finally, $t'=t-t_0$ where 
\be
t_0\=-4m^2\frac{\xi^2}{1-\xi^2}
\ee
is the minimal value of $-t$ allowed in the process in question. Since we
only consider small values of the skewness $-t_0$ is very 
small and the difference between $t'$ and $t$ is tiny.

An interesting property of the helicity amplitudes can be inferred from
\req{eq:TL}. With the help of parity conservation one sees that part of the 
amplitudes \req{eq:TL} behave like those for the exchange of a
particle with either natural ($N$) or unnatural parity ($U$)
\ba
{\cal M}^{V N}_{-\mu'\nu',-\mu\nu} &=& \phantom{-} (-1)^{\mu'-\mu} 
                                   {\cal M}^{V N}_{\mu'\nu',\mu\nu}\,,\nn\\
{\cal M}^{V U}_{-\mu'\nu',-\mu\nu} &=& - (-1)^{\mu'-\mu} 
                                   {\cal M}^{V U}_{\mu'\nu',\mu\nu}\,.
\label{eq:symmetry}
\ea
Thus, the combinations $\bar{E}_T-\xi \widetilde{E}$ and $\widetilde{H}_T$ behave 
like natural parity exchange while $\widetilde{E}_T-\xi E_T$ behaves
like unnatural parity. Remarkably, the proton helicity-flip amplitudes in 
\req{eq:TL} cannot be splitted in natural and unnatural parity contributions 
completely. Such a behavior of the amplitude ${\cal M}_{0-,++}$ is known to hold 
for photoproduction of pions since the late sixties \ci{phillips} and was the 
reason for the introduction of Regge cuts.
According to \ci{GK5} the GPDs $H$ ($\widetilde{H}$) and $E$ ($\widetilde{E}$) 
also behave like (un)natural parity exchange. The $\gamma^*_T\to V_T$ amplitudes 
can therefore be written as
\be
{\cal M}^V_{+\pm,++}\={\cal M}^{VN}_{+\pm,++}+{\cal M}^{VU}_{+\pm,++}\,.
\ee
Other $\gamma^*_T\to V_T$ amplitudes are related to these amplitudes either by
the symmetry \req{eq:symmetry} or by parity invariance. The amplitude 
${\cal M}^{VU}_{+-,++}$ is fed by the $\xi\tilde{E}$ \ci{GK1}. Since we are interested
in small skewness and since it is no reason known why $\tilde{E}$ could be larger
than the other GPDs (with the exception of the pion-pole contribution which is
however irrelevant for vector-meson production) we neglect ${\cal M}^{VU}_{+-,++}$.

With regard to the fact that the GPD $\widetilde{E}_T$ is antisymmetric in $\xi$:
$\widetilde{E}_T(\xi)=-\widetilde{E}_T(-\xi)$, we neglect $\widetilde{E}_T$ and 
$E_T$ in \req{eq:TL} for small skewness. Moreover, we also neglect the amplitude 
$H^V_{0-,-+}$ in \req{eq:TL} since it proportional to $t/Q^2$ due to angular 
momentum conservation in contrast to the helicity non-flip amplitude $H^V_{0-,++}$ 
which is not forced to vanish for forward scattering by this conservation law. 
Finally, we disregard the GPD $\widetilde{H}_T$ in \req{eq:TL} by the admittedly 
weak argument that its contribution is proportional to $t/(4m^2)$. Taking all 
these simplifications into account the amplitudes given in \req{eq:TL} reduce to
\ba
{\cal M}^V_{0-,++} &=& e_0 \sum_a e_a {\cal C}_V^a \int dx
                                   H^V_{0-,++}(x,\xi,Q^2,t=0)H^a_T(x,\xi,t)\,,\nn\\
{\cal M}^V_{0+,\pm +} &=& \mp e_0\frac{\sqrt{-t^\prime}}{4m} \sum_a e_a {\cal C}_V^a
                 \int dx H^V_{0-,++}(x,\xi,Q^2,t=0) \bar{E}^a_T(x,\xi,t)\,,\nn\\
{\cal M}^V_{0-,-+} &=& 0\,.
\label{eq:TL-simple}
\ea   
Although the transversity GPDs are leading twist, the amplitudes given in \req{eq:TL} 
and \req{eq:TL-simple} are of twist-3 nature. Quark and antiquark forming the 
valence Fock state of the longitudinally polarized vector meson have the same 
helicity in $H^V_{0-,++}$, see Fig.\ \ref{fig:handbag}. This necessitates the use 
of twist-3 meson wave functions which will be discussed in Sect.\ \ref{sec:twist3}. 

We repeat that \req{eq:TL-simple} only refers to the quark transversity GPDs.
The contributions from their gluonic partners require the non-flip subprocess 
amplitude $H^V_{--,++}$, i.e.\ the amplitude with gluon as well as
photon-meson helicity flip (all helicities are either plus or minus 1). The
convolutions of $H^V_{--,++}$ and the gluonic transversity GPDs determine the 
$\gamma^*_T\to V_{-T}^{\phantom{*}}$ amplitudes ${\cal M}^V_{\mp\nu',\pm\nu}$. As is 
well-known from the SDMEs for $\rho^0$ and $\phi$ 
production (e.g.\ $r^1_{11}$) measured for instance by HERMES \ci{hermes} and H1 
\ci{h1}, these amplitudes are very small, compatible with zero within errors and 
usually neglected in analyses of vector-meson leptoproduction~\footnote
{As shown in \ci{ji98,bm00} the gluon transversity GPDs contribute to the 
$\gamma^*_T\to \gamma^{\phantom{*}}_{-T}$ DVCS amplitudes to NLO.}. 
We will do so here as well. Small $\gamma^*_T\to V_{-T}^{\phantom{*}}$ amplitudes 
are consistent with the assumption of small gluonic transversity GPDs. This
assumption is not in conflict with rather large quark transversity GPDs since
the quark and gluon transversity GPDs evolve independently with the scale
\ci{ji98,bel00}. The amplitudes for $\gamma^*_L\to V_T^{\phantom{*}}$ 
transitions will be neglected too. They are experimentally small \ci{hermes,h1} and 
strongly suppressed in the handbag approach.

\subsection{Calculation of the twist-3 subprocess amplitude}
\label{sec:twist3}
We begin with the discussion of the light-cone wave function for the valence 
Fock component of a helicity-zero vector meson that moves along the 3-direction
and for which quark and antiquark have the same helicity, see Fig.\ \ref{fig:handbag}. 
Obviously, this configuration requires one unit of orbital angular momentum 
projection $l_3$. Such a light-cone \wf{} has been given in \ci{ji-yuan03} recently
\ba
|V;q',\mu'=0,|l_3|=1 \rangle &=& \frac1{\sqrt{2}}
      \int \frac{d\tau d^2\vk}{16\pi^3} \Psi^{(2)}_V(\tau,k^2_\perp)
                          \frac1{m_V\sqrt{\tau\bar{\tau}}} \nn\\
         &\times&   \Big[k^-_{\perp}b^\dagger_+(\tau,\vk){d}^\dagger_+(\bar{\tau},-\vk) \nn\\
  &-& \phantom{[} k^+_{\perp}  b^\dagger_-(\tau,\vk)
                {d}^\dagger_-(\bar{\tau},-\vk)\Big]\mid 0\rangle \,,
\label{eq:ji-yuan}
\ea
Color and flavor factors are omitted for convenience. The quark fields, $b^\dagger$ 
and $d^\dagger$, depend on the momentum fractions $\tau$ and $\bar{\tau}\equiv 1-\tau$ of 
the meson's momentum, $q'$, and on the quark transverse momentum, $\vk$. The 
combinations of its 1- and 2-components
\be
k^{\pm}_\perp \= k^1_\perp \pm i k^2_\perp
\ee
represent one unit of $l_3$. Acting on the perturbative vacuum the quark fields create 
quark and antiquark momentum eigenstates
\ba
\mid q' (\tau, \vk); \lambda\rangle &=& b^\dagger_{q\lambda}(\tau, \vk)\mid 0\rangle\,,\nn\\
\mid \bar{q}' (\bar{\tau}, -\vk); \lambda\rangle &=& 
                       d^\dagger_{q\lambda}(\bar{\tau}, -\vk)\mid 0\rangle\,.
\ea
It has been shown in \ci{ji-yuan03} that the wave function \req{eq:ji-yuan} 
has the correct behavior under the parity operation for a helicity-zero $\rho$ meson.
In contrast to \ci{ji-yuan03} we divide by the meson mass in order to have a
scalar \wf{} $\Psi^{(2)}$ of the same dimension as the \wf{} $\Psi^{(1)}$ appearing
in the expression for the usual $l_3=0$ Fock component of the vector meson
\ba
|V;q^\prime,\mu'=0,l_z=0 \rangle&=& \frac1{\sqrt{2}}
  \int \frac{d\tau d^2\vk}{16\pi^3} \Psi^{(1)}_V(\tau,k^2_\perp)\frac1{\sqrt{\tau\bar{\tau}}}\nn\\
        &\times& \Big[b^\dagger_+(\tau,\vk){d}^\dagger_{-}(\bar{\tau},-\vk)\nn\\
  &+&  \phantom{[}  b^\dagger_-(\tau,\vk){d}^\dagger_+(\bar{\tau},-\vk)\Big] \,\mid0\rangle\,.
\label{eq:l30-state}
\ea
The states \req{eq:ji-yuan} and \req{eq:l30-state} respect covariant particle state 
normalization. Hence, the probabilities of the $\mid l_3\mid=1$ and $0$ Fock components 
are given by
\ba
\int \frac{d\tau d^2\vk}{16\pi^3} \frac{k^2_\perp}{m_V^2} \mid \Psi_V^{(2)}(\tau,k^2_\perp)\mid^2
                            &=& P_{\mid l_3\mid=1}\,, \nn\\
\int \frac{d\tau d^2\vk}{16\pi^3} \mid \Psi_V^{(1)}(\tau,k^2_\perp)\mid^2
                            &=& P_{l_3=0}\,,
\ea
with $ P_{\mid l_3\mid=1}+P_{l_3=0}\leq 1$.
The spin part of \req{eq:ji-yuan} is equivalent to the following expression
\be
\Gamma_{|l_3|=1}\=\frac{1}{\sqrt{2}m_V}
         \Big[q\sla'k\sla +m_Vk\sla - \frac{k^2_\perp}{2\tau\bar{\tau}}  
            +k^2_\perp\frac{\bar{\tau}-\tau}{2\tau\bar{\tau}m_V}  q\sla'
               + {\cal O}(k^3_\perp)\Big]\,.
\label{eq:cov-spin-wf}
\ee
for an incoming vector meson. The 4-vector $k$ is defined as 
\be
k\=[0,0,\vk]\,,
\ee
in light-cone coordinates. This spin \wf{} can be transformed to the frame
we are working by a transverse boost. The equivalence of \req{eq:cov-spin-wf}
and the spin part of \req{eq:ji-yuan} can readily be derived. Representing the
parton states in \req{eq:ji-yuan} by Dirac spinors in the rest frame, one sees
\be
k^-_\perp u_+(0)\bar{v}_+(0) - k^+_\perp u_-(0)\bar{v}_-(0)\= \frac12 (1+\gamma^0)k\sla\,.
\ee
A boost of this expression to the frame where the meson moves rapidly along the 
3-axis leads to
\ba
k^-_\perp u_+(\tau,\vk)\bar{v}_+(\bar{\tau},-\vk) &-& k^+_\perp u_-(\tau,\vk)\bar{v}_-(\bar{\tau},-\vk)
\nn \\ 
 &\sim & (p\sla_1+m_1)(q\sla'+m_V)k\sla (-p\sla_2+m_2)
\label{eq:expression1}
\ea
with the quark and antiquark momenta being defined as
\be
p_1\=[\tau q'{}^{+},\frac{\tau^2m_V^2+k^2_\perp}{2\tau q^{'+}},\vk]\,, \qquad
p_2\=[\bar{\tau} q'{}^{+},\frac{\bar{\tau}^2m_V^2+k^2_\perp}{2\bar{\tau} q^{'+}},-\vk]\,.
\ee
The quark and antiquark masses are taken as $m_1=\tau m_V$ and $m_2=\bar{\tau} m_V$.
This guarantees that $q'=p_1+p_2$ up to corrections of order $k^2_\perp$. From \req{eq:expression1}
one easily derives \req{eq:cov-spin-wf}.

By counting the numbers of $\gamma$ matrices in the Feynman expression for this amplitude
(including the two from the proton matrix element for parton helicity flip) one sees that 
only the first and the third term of the spin \wf{} \req{eq:cov-spin-wf} contribute to the
parton helicity-flip amplitude. The first term, $q\sla' k\sla$, leads to a contribution 
of order $t/Q^2$ and is consequently neglected. Hence, the subprocess amplitude 
$H^V_{0-,++}$ is generated by the third term. Performing the LO calculation of $H^V_{0-,++}$
from that term and the set of Feynman graphs of which an example is shown in Fig.\ 
\ref{fig:handbag}, we obtain 
\ba
H^{V}_{0-++}&=&32\pi \frac{m_V\xi}{Q^2} \frac{C_F}{\sqrt{N_c}} \int d\tau 
             \int \frac{dk^2_\perp}{16\pi^2}
         \frac{k^2_\perp}{2\tau\bar{\tau}m_V^2} \Psi_V^{(2)}(\tau,k^2_\perp) \nn\\ 
        &\times& \als(\mu_r) 
     \left(\frac1{x-\xi+i\eps}\frac1{\bar{\tau}(x-\xi)-2\xi k^2_\perp/Q^2+i\eps}\right.\nn\\
         && \left.  + \frac1{x+\xi-i\eps}\frac1{\tau(x+\xi)+2\xi k^2_\perp/Q^2-i\eps}\right)\,.
\label{eq:subampl}
\ea
The number of colors is denoted by $N_c$, $C_F=4/3$ and $\mu_R$ is an appropriate 
renormalization scale (see below). Eq.\ \req{eq:subampl} holds for unflavored vector 
mesons. As we already mentioned for flavored mesons built up by a quark $q_a$ and an 
antiquark $\bar{q}_b$, the corresponding quark charges $e_a$ and $e_b$ multiply the 
first and second term of \req{eq:subampl}, respectively. Following \ci{li-sterman} 
we only retain $k_\perp^2$  in the denominators of the parton propagators. There the 
parton transverse momentum plays a decisive role since it competes with terms 
$\propto \tau (\bar{\tau})Q^2$ which become small in the end-point regions where either 
$\tau$ or $\bar{\tau}$ tends to zero.

The \da{} associated with the third term of the \wf{} \req{eq:cov-spin-wf}, reads
\be
\int \frac{dk^2_\perp}{16\pi^2}\frac{k^2_\perp}{2\tau\bar{\tau}m^2_V}\Psi^{(2)}_V(\tau,k^2_\perp)\=
              \frac{f^T_V}{2\sqrt{2N_c}} h_{\| V}^{(s)}(\tau)\,.
\ee
According to \ci{ball-braun98}, the twist-3 chiral-odd \da{} $h_{\|}^{(s)}$ is defined by the 
meson-vacuum matrix element~\footnote{
A second twist-3 helicity-flip \da{}, $h_{\| V}^{(t)}$, \ci{ball-braun98} is associated with 
the  $q\sla' k\sla$ -term of the $|l_3|=1$ wave function.} 
\be  
\langle 0| \bar{q}(z)q(-z)|V;q^\prime,\mu'=0\rangle
\ee
(a path-ordered gauge factor along the straight line connecting the points $z$ and 
$-z$ is understood). This distribution amplitude comes along with the 
tensor decay constant $f_V^T$ of the vector meson. The latter depends on the
factorization scale $\mu_F$ to be specified below
\be
f^T_V(\mu_F)\=f^T_V(\mu_0)\left(\frac{\als(\mu_F)}{\als(\mu_0)}\right)^{4/27}\,.
\ee
For the tensor decay constant we use the QCD sum rule estimate give in \ci{ball-braun96}.
According to this work it amounts to about 0.8 times the usual decay constant of a 
longitudinally polarized $l_3=0$ vector meson at the scale $\mu_0=1\,\gev$. As a 
consequence of the nature of the \wf{} $\Psi_V^{(2)}$ the subprocess amplitude 
$H^V_{0-,++}$ is of twist-3 accuracy and is parametrically suppressed by $m_V/Q$ as 
compared to the leading-twist amplitudes $H^V_{0+,0+}$.

In the modified perturbative approach we are using, the amplitude \req{eq:subampl} is
Fourier transformed from the $\vk$-space to the canonically conjugated impact parameter 
space $\vbs$, for details see \ci{GK3}. The obtained $\vbs$-space expression is multiplied
with the Sudakov factor, $\exp{[-S(\tau,\vbs,Q^2)]}$, representing gluon radiation 
calculated to next-to-leading-log accuracy using resummation techniques and having 
recourse to the renormalization group \ci{li-sterman}. The impact parameter $\vbs$ which
is the quark-antiquark separation, acts as an infrared cut-off. Radiative gluons with
wave lengths between the infrared cut-off and a lower limit (related to the hard scale $Q^2$)
yield suppression; softer gluons are part of the meson \wf, while harder ones are an 
explicit part of the subprocess amplitude. Consequently, the factorization scale is
given by the quark-antiquark separation $\mu_F=1/b$. The renormalization scale, $\mu_R$,
is taken to be the largest mass scale appearing in the subprocess amplitude, i.e.\
$\mu_R={\rm max}(\tau Q,\bar{\tau}Q,1/b)$. For $\Lambda_{\rm QCD}$ a value of $220\,\mev$ 
is used in the Sudakov factor and in the evaluation of $\als$ from the one-loop expression.

\subsection{Parametrization of the GPDs}
In order to evaluate the convolutions in \req{eq:TL-simple} and the analogous ones for the
other amplitudes we need the GPDs. We adopt for them the parametrizations proposed in our 
previous work \ci{GK5,GK6,GK3}. The GPDs are constructed from the zero-skewness GPDs with 
the help of the double distribution ansatz \ci{muel98} 
\be
K^i(x,\xi,t)\=\int_{-1}^1 d\rho\, \int_{-1+\mid\rho\mid}^{1-\mid\rho\mid}d\eta \delta(\rho+\xi\eta-x)
              K^i(\rho,\xi=0,t) w_i(\rho,\eta)\,,
\ee
where $K$ is a GPD and $i$ stands for gluon, sea or valence quarks. A possible 
$D$ term \ci{pol-weiss} is neglected. For the weight function $w$ that generates the skewness
dependence we use \ci{mus99}
\be
w_i(\rho,\eta)\=\frac{\Gamma(2n_i+2)}{2^{2n_i+1}\Gamma^2(n_i+1)}\,
               \frac{[(1-\mid\rho\mid)^2-\eta^2]^{n_i}}{(1-\mid\rho\mid)^{2n_i+1}}\,.
\ee
For the parameter $n_i$ a value of 2 is taken for the gluon and sea-quark helicity non-flip 
GPDs and 1 in all other cases. The zero-skewness GPDs are parametrized as 
\be
K^i(\rho,\xi=0,t)\=k^i(\rho)\exp{[t p_{ki}(\rho)]}\,,
\label{eq:zero-skewness}
\ee
where $k^i$ is the forward ($t=0$) limit of the zero-skewness GPD which for $H$, 
$\widetilde{H}$ and $H_T$ are the unpolarized, polarized and transversity PDFs, respectively. 
For the other GPDs the forward limits are parametrized like the PDFs
\be
k^i(\rho)\=N_{ki}\rho^{-\alpha_{ki}}(1-\rho)^{\beta_{ki}}\,.
\label{eq:forward-limit}
\ee
The profile function $p_{ki}$ in \req{eq:zero-skewness} is parametrized in a Regge-like 
manner
\be
p_{ki}(\rho)\=-\alpha^\prime_{ki}\ln{(\rho)} + b_{ki}\,,
\ee
where $\alpha^\prime_{ki}$ represents the slope of a Regge trajectory and $b_{ki}$ parametrizes
the $t$ dependence of its residue.

The best determined GPD is $H$ since it controls the cross sections for leptoproduction
of flavor-neutral vector mesons. The values of the parameters which specify $H$, are obtained 
from fits to the cross section data at small skewness and can be found in \ci{GK3}. The 
GPDs $\widetilde{H}$ and $\widetilde{E}$ play no role in the observables we are going to 
discuss below. The GPD $E$ for valence quarks, on the other hand, is of importance for 
some of the observables of interest. The values of its parameters are given in \ci{GK3,GK4}.
This parametrization of $E$ for valence quarks at zero skewness is in agreement with the 
findings of an analysis of the nucleon form factors in terms of GPDs \ci{DFJK4}. According to 
this analysis the second moments of $E$ for $u$ and $d$ valence quarks at $t=0$ have about the 
same magnitude but opposite sign. Due to a sum rule for the second moments of $E$ at 
$\xi=t=0$ \ci{teryaev,diehl} the respective moments for the gluon and sea quarks cancel each 
other to a large extent. Since, for our parametrization, the zero-skewness GPDs 
have no nodes  except at the end points $x=0$ and $1$, this cancellation approximately happens 
for other moments too. It even approximately occurs for the convolutions with the subprocess 
amplitudes. For this reason we do not consider $E$ for gluons and sea quarks in this work. In 
passing we note that the set of helicity non-flip GPDs proposed in \ci{GK3,GK4} has been 
examined in a 
calculation of DVCS to leading-twist accuracy and leading-order of perturbative QCD \ci{KMS}. 
The results are found to be in satisfactory agreement with all small skewness data. Recently 
the form factor analysis from 2004 \ci{DFJK4} has been updated \ci{DK13}. All the new data
on the nucleon form factors are taken into account in the update as well as more recent parton 
distributions \ci{ABM12}. The zero-skewness valence-quark GPDs $H$ and $E$ obtained in this 
analysis do not differ much from those proposed in the 2004 analysis at low $-t$. We checked 
that the use of these new valence-quark GPDs do not alter our results perceptibly.  
  
The only available small-skewness data which provide clear evidence for strong contributions 
from transversely polarized virtual photons and therefore information on the transversity GPDs,
are the $\pi^+$ electroproduction cross section \ci{hermes07} and the asymmetries measured 
with a transversely polarized target \ci{hermes09}. However the $\pi^+$ data provide only 
information on the combination $H_T^u-H_T^d$. The forward limit of $H_T$ is the transversity 
distribution, $\delta(x)$, which has been determined by Anselmino {\it et al.} \ci{anselmino08} 
in an analysis of the data on the azimuthal asymmetry in semi-inclusive deep inelastic 
lepton-nucleon scattering and in inclusive two-hadron production in electron-positron 
annihilation. The moments of the transversity distributions proposed in \ci{anselmino08}, i.e.\ 
the lowest moments of $H_T$ at $t'=0$,  are about $40\%$ smaller than a lattice QCD result 
\ci{lattice05}, they are also substantially smaller than model results (cf.\ \ci{anselmino08} 
and references therein). Also the analysis of $\pi^0$ leptoproduction performed in \ci{GK6}, 
suggest larger moments of $H_T$. In order to surmount this difficulty we leave  unchanged 
the parametrization of the transversity distributions given in \ci{anselmino08,GK6} but adjust 
their normalizations to the lattice QCD moments of \ci{lattice05}. The other transversity GPD, 
$\bar{E}_T$, is only constrained by lattice QCD results \ci{lattice06}, its contribution to 
$\pi^+$ production is very small. The values of the parameters for the valence quark GPDs 
$H_T$ and $\bar{E}_T$ proposed in \ci{GK6}, are quoted in Tab.\ \ref{tab:1}. Given the 
uncertainties of the present lattice QCD results \ci{lattice12} we consider these 
parametrizations as rough estimates which only allow explorative studies of transversity 
effects in exclusive meson leptoproduction. In other words, we only achieve estimates of 
various observables. For this reason we do not attempt an error assessment of our results; 
this is beyond feasibility at present. Evolution of the transversity GPDs is not taken into 
account, all pertinent experimental data cover only a very limited range of $Q^2$.

The last item we have to specify are the sea-quark transversity GPDs. A flavor symmetric 
sea is assumed with the parameters quoted in Tab.\ \ref{tab:1}. These parameters are 
adjusted to the data discussed below.
\begin{table*}[h]
\renewcommand{\arraystretch}{1.4} 
\begin{center}
\begin{tabular}{| c || c | c | c | c | c |}
\hline   
GPD & $\alpha_{ki}$ & $\beta_{ki}$ & $\alpha^\prime_{ki} [\gev^{-2}]$ & $b_{ki} [\gev^{-2}]$ & $N_{ki}$ 
\\[0.2em]  
\hline
$H_T^{u_v}$       & -     & 5 & 0.45 & 0.3 & 1.1 \\[0.2em] 
$H_T^{d_v}$       & -     & 5 & 0.45 & 0.3 & -0.3 \\[0.2em]
$H_T^s$          & 0.6   & 7 & 0.45 & 0.5 & -0.17  \\[0.2em]
\hline
$\bar{E}_T^{u_v}$ & 0.3   & 4 & 0.45 & 0.5 & 6.83 \\[0.2em]
$\bar{E}_T^{d_v}$ & 0.3   & 5 & 0.45 & 0.5 & 5.05 \\[0.2em]
$\bar{E}_T^s$    & 0.6   & 7 & 0.45 & 0.5 & -0.10 \\[0.2em]
\hline
\end{tabular}
\end{center}
\caption{Parameters for the transversity GPDs at a scale of $2\,\gev$.}
\label{tab:1}
\renewcommand{\arraystretch}{1.0}   
\end{table*} 

The $l_3=0$ \wf s for the vector mesons are specified in \ci{GK3,GK4}. Basically they are
simple Gaussians in $\vk$. This type of \wf{} is also used for the scalar $\mid l_3\mid=1$
\wf{}
\be
\Psi^{(2)}_V(\tau,k^2_\perp)\= 16\pi^2\sqrt{2N_c}f_V^Tm^2_Va_{VT}^4
                                              \exp{[-a_{VT}^2k^2_\perp/(\tau\bar{\tau})]}\,.
\label{eq:wf}
\ee  
Its associated \da{} is just the asymptotic form for mesons
\be
h_{\| V}^{(s)}=6\tau\bar{\tau}\,.
\ee
In principle this is the leading term of a Gegenbauer series \ci{ball-braun98}. We however 
disregard all higher Gegenbauer terms except of the $C_1^{3/2}$-term for the $K^{*0}$ meson 
for which we take a value of 0.1 for its coefficient. As discussed in \ci{kroll10} the 
higher Gegenbauer terms are strongly suppressed in the modified perturbative approach.  

The \wf{} \req{eq:wf} leads to the probability of the $\mid l_3\mid=1$ Fock component 
\be
P_{\mid l_3\mid=1}\=\frac4{15}\, \pi N_c \big(f_V^Tm_Va^2_{VT}\big)^2
\ee 
and the r.m.s.\ $k_\perp$ is 
\be
\langle k^2_\perp \rangle \= \frac3{14}\,a_{VT}^{-2}\,.
\ee
With $a_{\rho T}\simeq 1\,\gev$ and $f_\rho^T=167\,\mev$  (see \ci{GK3}) one 
finds the plausible values $P_{\mid l_3\mid=1}=0.13$ and $\langle k^2_\perp\rangle^{1/2}=0.46\,\gev$. 

\section{Results}
\label{sec:results}

\subsection{Spin density matrix elements}
\label{sec:sdmes}
The $\gamma^*_T\to V_L^{\phantom{*}}$ amplitudes can be probed by some of the SDMEs. 
Using the simplifications discussed in Sect.\ \ref{sec:handbag}, one finds for
the relevant SDMEs \ci{schilling}
\ba
r_{00}^1(V) \sigma_0^V&=& -\mid{\cal M}_{0+++}^V\mid^2\,, \nn\\ 
r_{00}^5(V) \sigma_0^V&=& \sqrt{2}\,{\rm Re}
\Big[{\cal M}_{0+++}^{V*} {\cal M}_{0+0+}^V + \frac12{\cal M}_{0-++}^{V*} {\cal M}_{0-0+}^V\Big]\,, \nn\\
{\rm Re}\,r^{04}_{10}(V)\sigma_0^V &=& -{\rm Re}\, r^{1}_{10}(V)\sigma_0^V \={\rm Im}\, 
                                 r^2_{10}(V)\sigma_0^V \nn\\
          &=& \frac12\,{\rm Re}\Big[
        {\cal M}_{0+++}^{V*} {\cal M}_{++++}^{VN} 
       + \frac12 {\cal M}_{0-++}^{V*} {\cal M}_{+-++}^{VN}\Big]\,,
\label{eq:sdme}
\ea
where 
\ba  
\sigma_0^V &=& \mid{\cal M}_{++,+ +}^V\mid^2 + \mid{\cal M}_{+-,+ +}^V\mid^2  
             + \mid{\cal M}_{0+,+ +}^V\mid^2 + \frac12\mid{\cal M}_{0-,+ +}^V\mid^2\nn\\ 
         &+& \varepsilon \Big[\mid{\cal M}_{0+,0 +}^V\mid^2
            + \mid{\cal M}_{0-,0 +}^V\mid^2\Big]\,.
\label{eq:norm}
\ea
The ratio of the longitudinal and transverse photon flux is denoted by $\varepsilon$.
Up to a phase space factor $\sigma_0^V$ is the unseparated cross section 
$d\sigma = d\sigma_T +\varepsilon d\sigma_L$. The contribution from the $\gamma^*_T\to V_L^{\phantom{*}}$ 
amplitudes to the transverse cross section for $\rho^0$ production is
negligibly small, it amounts to only $2-3\%$.

\begin{figure}[t] \begin{center}
\includegraphics[width=0.45\tw]{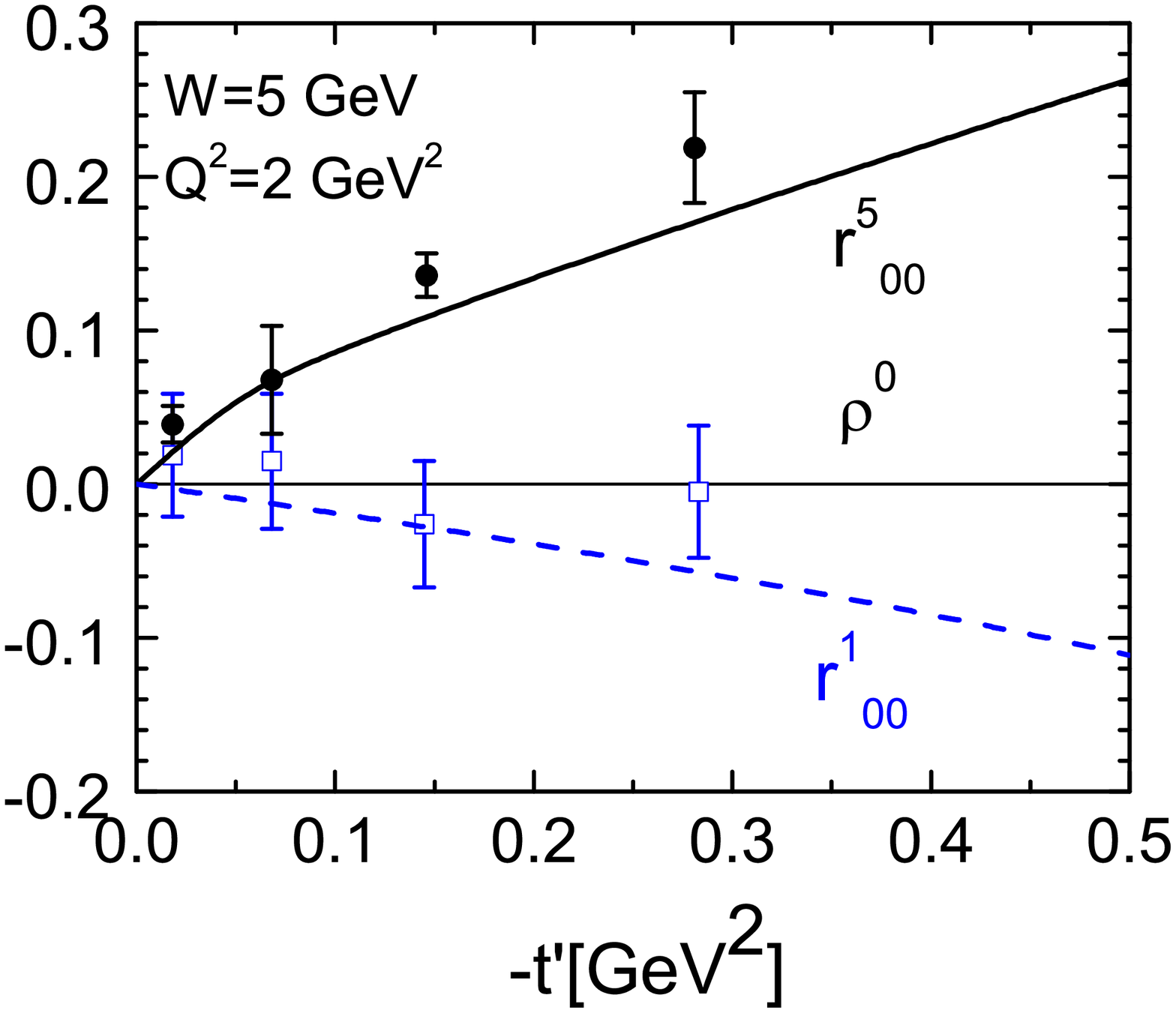}
\includegraphics[width=0.45\tw]{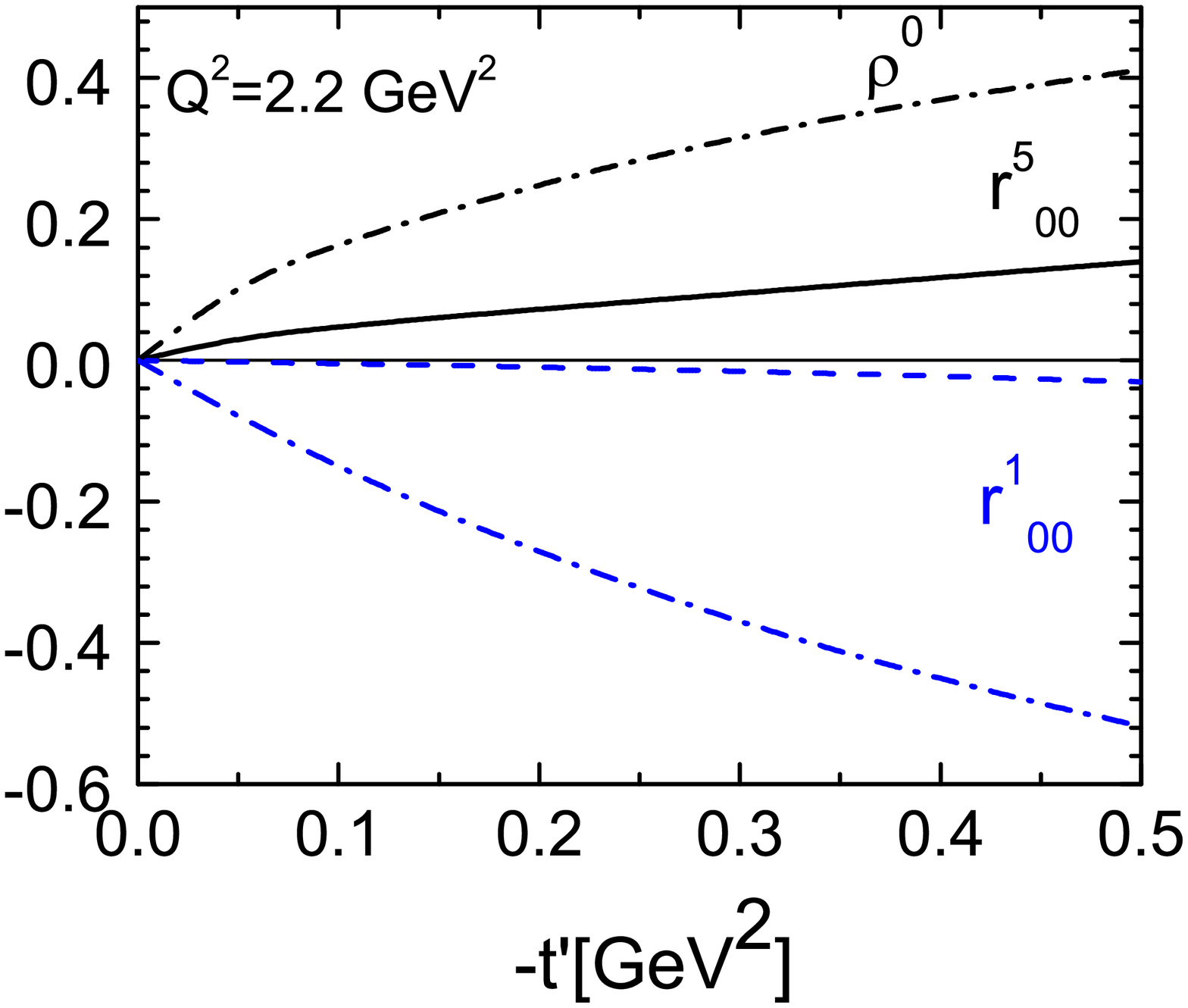}
\caption{Left: Handbag results for the SDMEs $r^5_{00}$ (solid line) and $r^1_{00}$ (dashed line) 
for $\rho^0$ production. Data taken from HERMES \ci{hermes}. Right: Predictions for $r^5_{00}$
and  $r^1_{00}$ at $W=8.1\,\gev$ (solid and dashed line, respectively) and $W=3\,\gev$ (dash-dotted
lines).}
\label{fig:sdme1}
\end{center}
\end{figure}
A particularly interesting SDME is $r^1_{00}$. It measures the absolute value of the 
amplitude ${\cal M}_{0+++}$ which is fed by the GPD $\bar{E}_T$ in the combination
$e_u\bar{E}_T^u-e_d\bar{E}_T^d$ for $\rho^0$ production, see \req{eq:flavor} and \req{eq:TL-simple}. 
Since both, $\bar{E}_T^u$ and $\bar{E}_T^d$, have the same sign and almost the same strength 
this amplitude is rather large. The signs of these GPDs are fixed by the lattice QCD results 
\ci{lattice06}. In fact, for the tensor anomalous magnetic moment of the nucleon
which represents the lowest moment of $\bar{E}_T$ at $t=0$, $\kappa_T^u\simeq \kappa_T^d > 0$ 
is found in \ci{lattice06}. Models support this result \ci{burkardt06,pasquini05}.

The SDME $r^5_{00}$ is more complicated. It measures the real part of a combination of 
two interference terms; in terms of GPDs
\be
r^5_{00} \sim {\rm Re} \Big[\langle \bar{E}_T\rangle^*_{LT}\langle H\rangle_{LL}^{\phantom{*}}
                         +\frac12\langle H_T\rangle^*_{LT}\langle E\rangle_{LL}^{\phantom{*}}\Big]
\label{eq:r500}
\ee
where $\langle K\rangle_{XY}$ denotes the convolution of the GPD $K$ with the subprocess 
amplitude for a $\gamma^*_Y\to V_X^{\phantom{*}}$ transition ($X, Y$  label longitudinal or 
transverse polarization ). I.e.\ $r^5_{00}$ is related to interference terms
of amplitudes fed by transversity GPDs with leading $\gamma^*_L\to V_L^{\phantom{*}}$ amplitudes. 
The first term in \req{eq:r500} dominates by far since $\langle H \rangle_{LL}$ is much larger 
than $\langle E \rangle_{LL}$ while both the transversity contributions are of roughly the same 
strength. Thus, $r^5_{00}$ essentially probes $\bar{E}_T$, too. As is to be seen from Fig.\ 
\ref{fig:sdme1} we achieve fair agreement  between the HERMES data on $\rho^0$ production
\ci{hermes} and our handbag results for $r^1_{00}$ and $r^5_{00}$. A point worth mentioning is 
that $r_{00}^5\propto \sqrt{-t'}$ and $r_{00}^1\propto t'$ for $t'\to 0$ as a consequence of 
angular momentum conservation.
 
\begin{figure}[t]
  \begin{center}
    \includegraphics[width=0.45\tw]{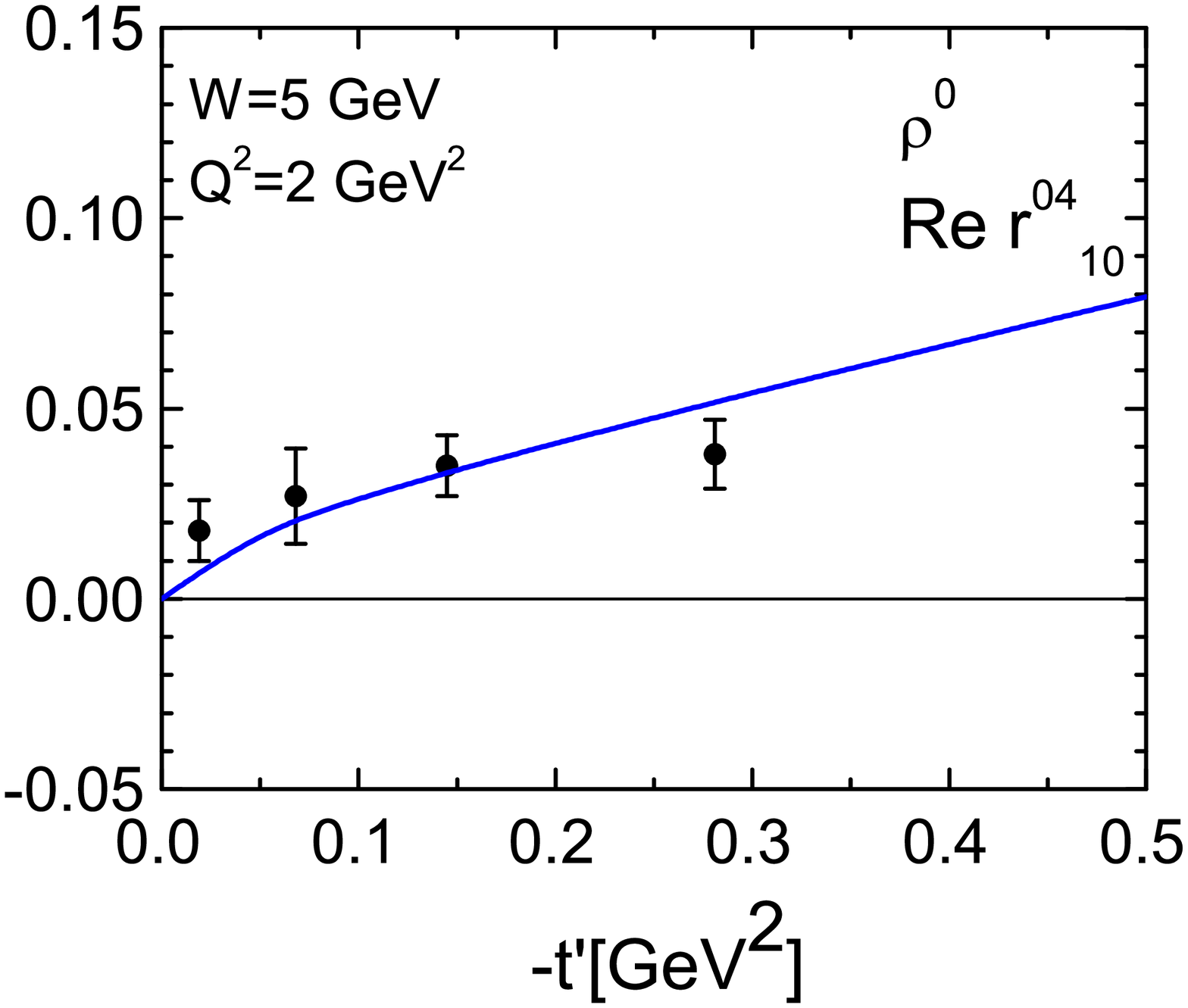}\hspace*{0.05\tw}
    \includegraphics[width=0.45\tw]{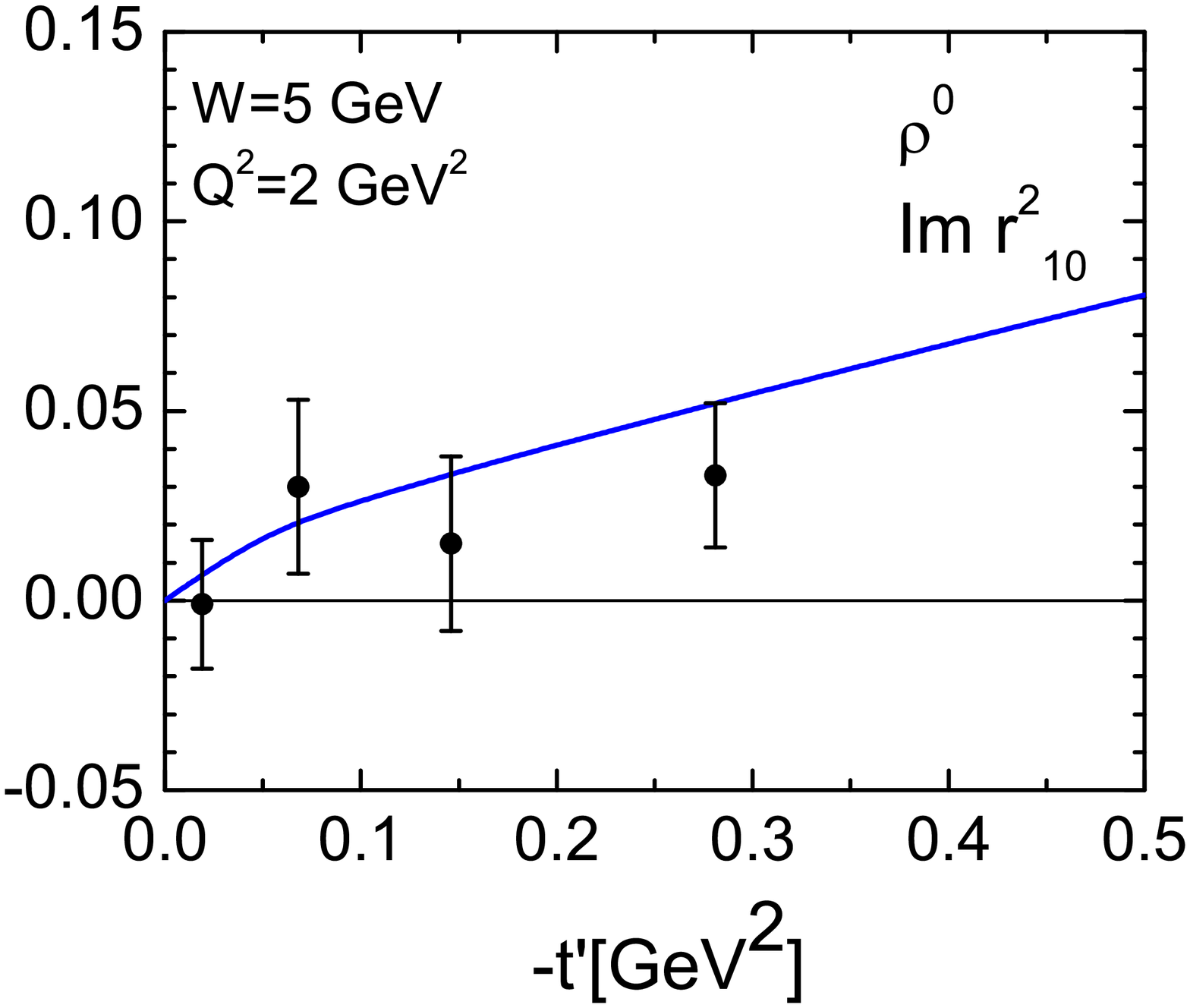}
  \end{center}
  \caption{The SDMEs ${\rm Re}\, r^{04}_{10}$ and ${\rm Im}\, r^2_{10}$. Data taken from HERMES 
    \ci{hermes}. The solid lines represent our results.}
  \label{fig:sdme2}
\end{figure}
Also for the SDMEs ${\rm Re}\,r^1_{00}$, ${\rm Re}\, r^{04}_{10}$ and ${\rm Im}\, r^2_{10}$ we 
find good agreement with the data on $\rho^0$ production, see Fig.\ \ref{fig:sdme2} for the
latter two SDMEs. Within the handbag approach the three SDMEs are equal (up to a sign) and 
probe a similar combination of interference terms as $r_{00}^5$. The difference is that for 
these SDMEs $H$ and $E$ are convoluted with the subprocess amplitude for 
$\gamma^*_T\to V_T^{\phantom{*}}$ transitions. For these SDMEs the $\bar{E}_T$ term is also 
dominant. The contribution from $\bar{E}_T$ for sea quarks is less than $5\%$ for all 
SDMEs, the valence quarks dominate.

These results are an addendum to our previous study of SDMEs \ci{GK3}. In summary we achieve 
a fair description of all SDMEs within the handbag approach now. An exception is the relative 
phase between the amplitudes for $\gamma^*_T\to \rho^0_T$ and $\gamma^*_L\to \rho^0_L$ 
transitions which is too small in the handbag approach as compared to experiment \ci{hermes}.
It would be of interest to probe the transversity contributions to the SDMEs also at other 
energies. As an example we show in Fig.\ \ref{fig:sdme1} $r_{00}^5$ and $r_{00}^1$ at the 
COMPASS energy of $8.1\,\gev$ and at $3\,\gev$ which is typical of the upgraded JLab.

\subsection{Transversely polarized target asymmetries}
\label{sec:asymmetries}
There are the following non-zero modulations of the transverse target spin asymmetry $A_{UT}$ 
\ba
A_{UT}^{\sin(\phi-\phi_s)}(V) \sigma_0^V &=&-2\, {\rm Im}\Big[\varepsilon
                         {\cal M}_{0-,0+}^{V*}{\cal M}_{0+,0+}^V \nn\\[0.2em]
                && +{\cal M}_{+-,++}^{VN*} {\cal M}_{++,++}^{VN}
                + \frac12\,{\cal M}_{0-,++}^{V*} {\cal M}_{0+,++}^V\Big]\,,\nn\\[0.2em]
A_{UT}^{\sin(\phi_s)}(V) \sigma_0^V &=& \sqrt{\varepsilon(1+\varepsilon)}\, {\rm Im}
            \Big[{\cal M}_{0+++}^{V*} {\cal M}_{0-0+}^V 
                -{\cal M}_{0-++}^{V*} {\cal M}_{0+0+}^V\Big]\,, \nn\\[0.2em]
A_{UT}^{\sin(\phi+\phi_s)}(V) \sigma_0^V &=&\varepsilon\, {\rm Im} \Big[{\cal
  M}_{0-,++}^{V*}{\cal M}_{0+,++}^V \Big]\,,\nn\\[0.2em]
A_{UT}^{\sin(2\phi-\phi_s)}(V) \sigma_0^V &=&-\sqrt{\varepsilon(1+\varepsilon)}\,
               {\rm Im}\Big[{\cal M}_{0+,++}^{V*} {\cal M}_{0-,0+}^V \Big]\,.
\label{eq:aut}
\ea
which can easily be derived from expressions given in \ci{diehl-sapeta}~\footnote{
The angle between the directions of the virtual photon and the incoming lepton is negligibly
small for the kinematics of interest in this work.}.
Here, $\phi$ is the azimuthal angle between the lepton and the hadron plane and $\phi_s$ 
specifies the orientation of the target spin vector with respect to the lepton plane. It 
is to be stressed that the COMPASS collaboration \ci{compass-aut} which has measured these 
modulations recently, took out the $\varepsilon$-dependent prefactors 
$\sqrt{\varepsilon (1\pm \varepsilon)}$ and $\varepsilon$ (for the $\sin(\phi + \phi_s)$ 
modulation) in their definition of the asymmetries ($\varepsilon\simeq 0.8$ for HERMES
and $\simeq 0.96$ for COMPASS kinematics). 

\begin{figure}[t]
\begin{center}
\includegraphics[width=0.36\tw]{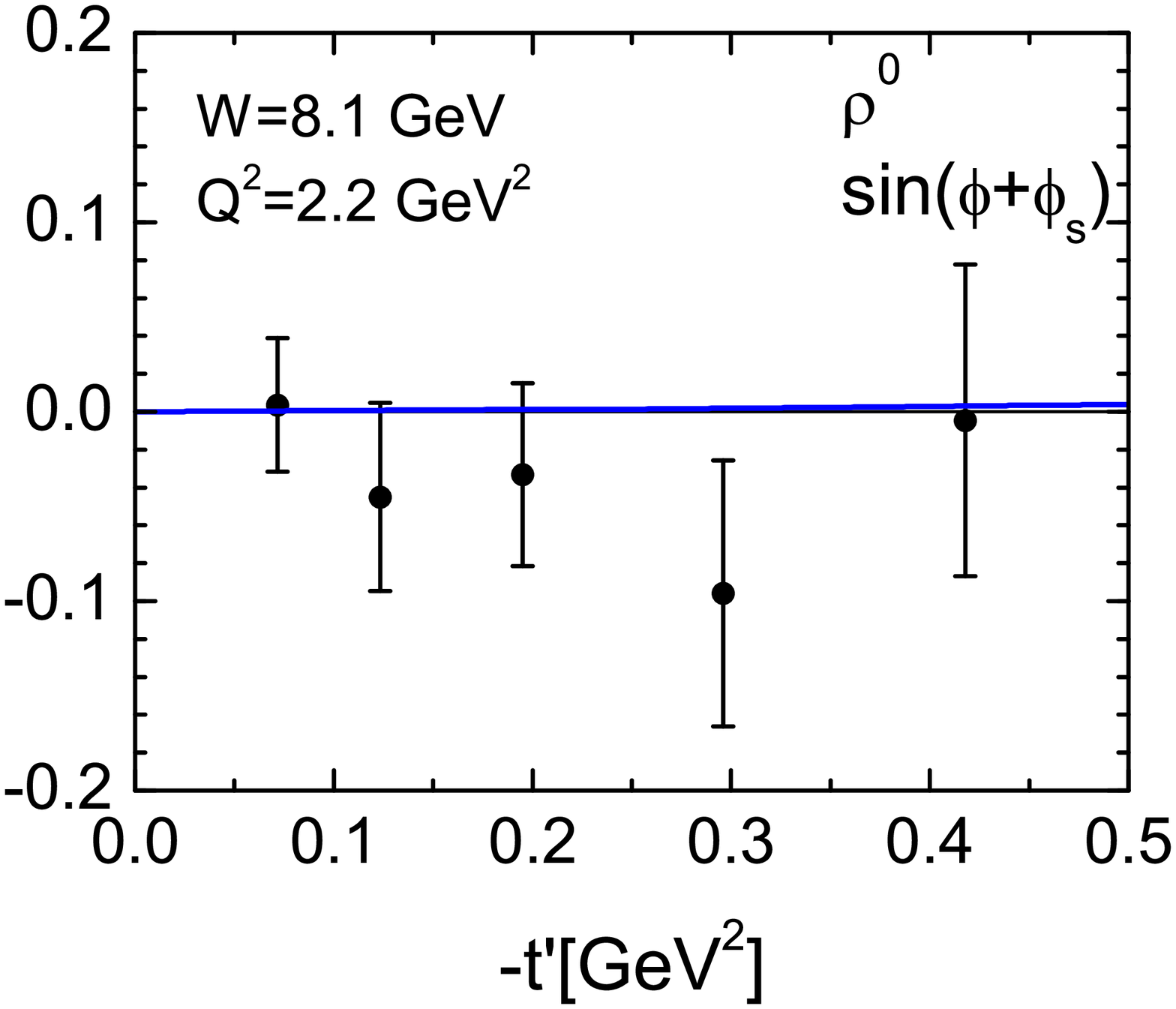}\hspace*{0.05\tw}
\includegraphics[width=0.36\tw]{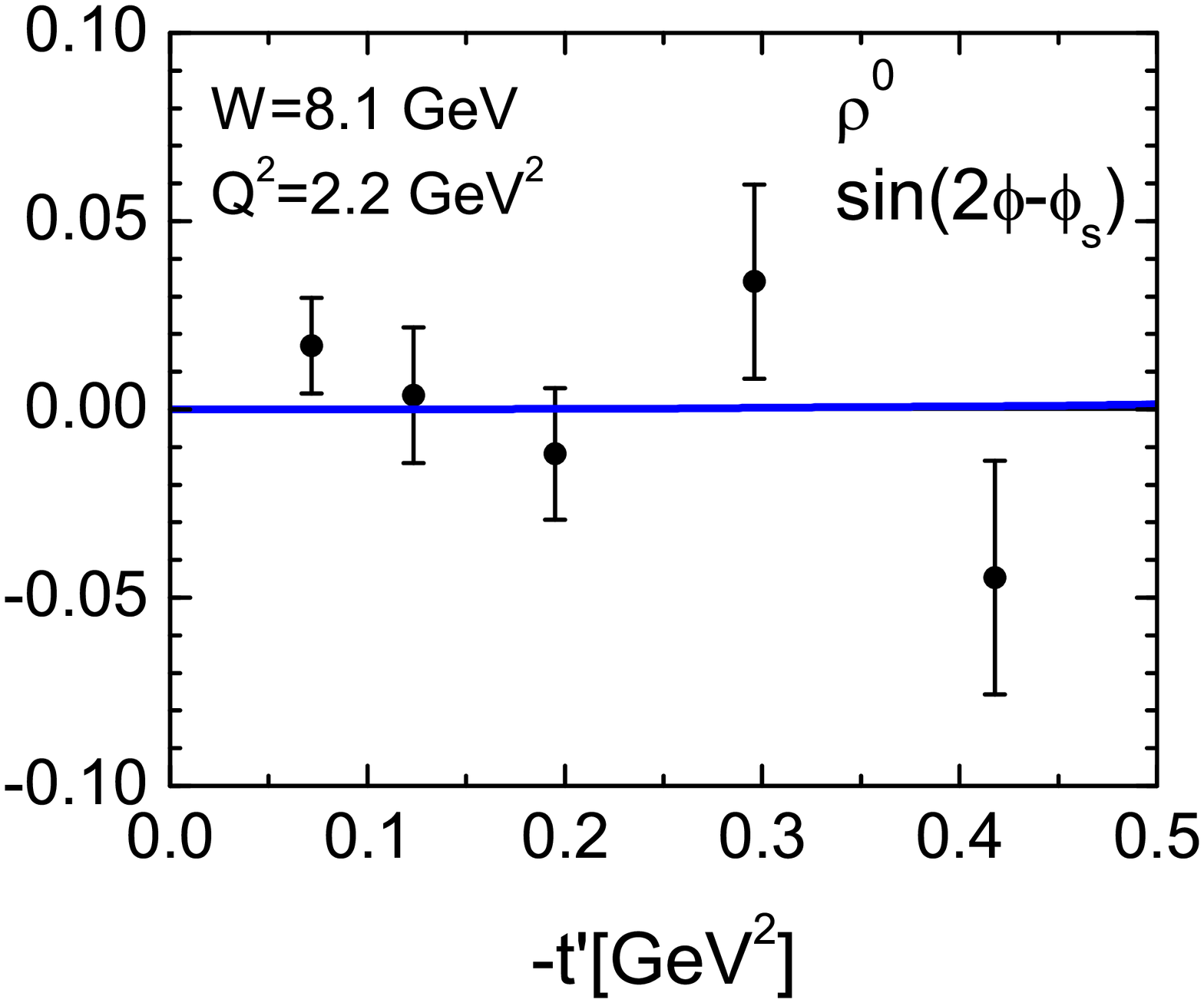}
\end{center}

\vspace*{-0.05\tw}
\caption{The $\sin{(\phi + \phi_s)}$ and $\sin{(2\phi -\phi_s)}$ modulations of $A_{UT}$ 
for $\rho^0$ leptoproduction divided by $\varepsilon$ and $\sqrt{\varepsilon(1+\varepsilon)}$, 
respectively. The results from the handbag approach are represented by solid lines. Data are 
taken from COMPASS \ci{compass-aut}.}
\label{fig:aut-compass}
\begin{center}
\includegraphics[width=0.36\tw]{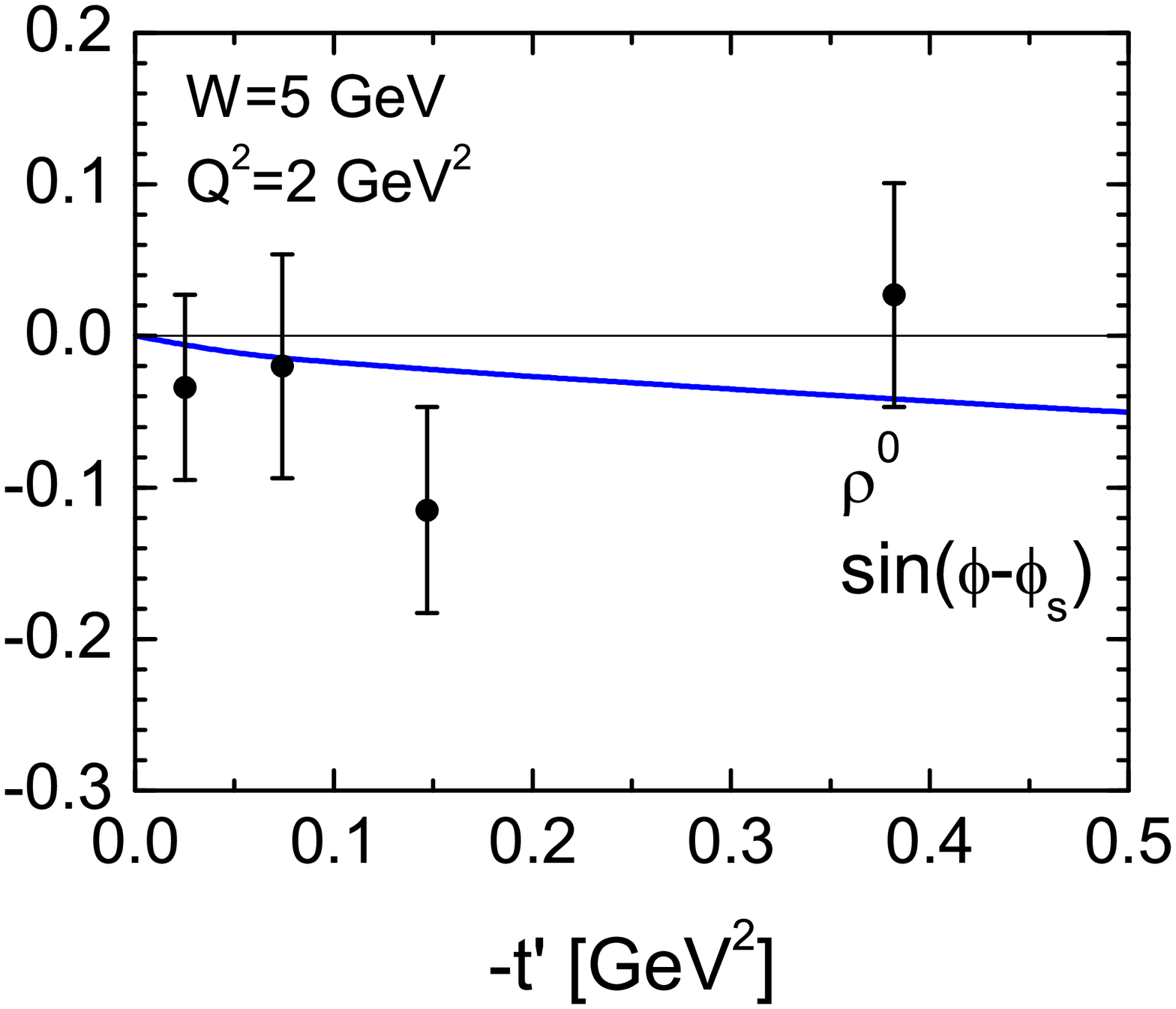}\hspace*{0.05\tw}
\includegraphics[width=0.37\tw]{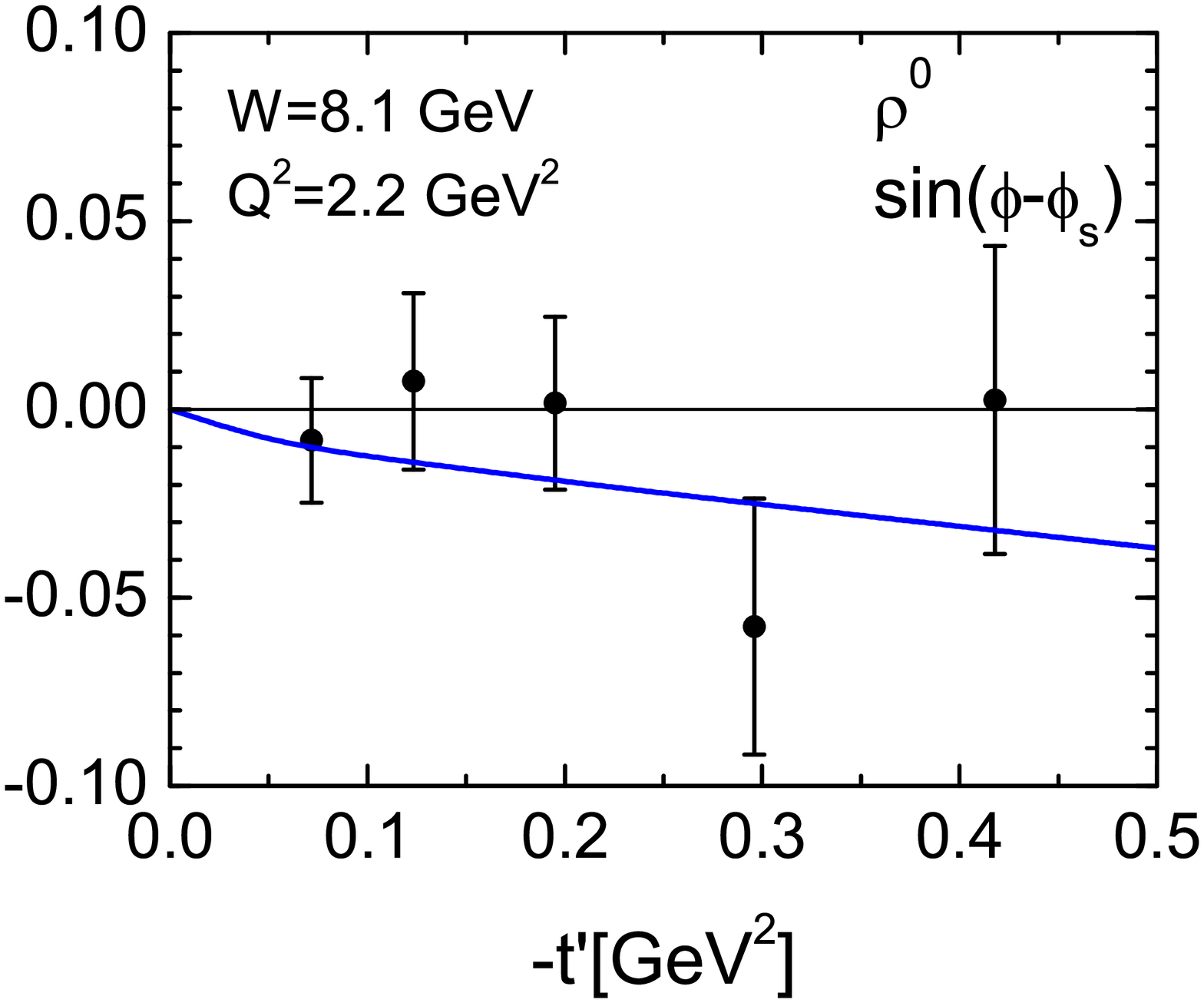}
\end{center}

\vspace*{-0.05\tw}
\caption{The $\sin{(\phi-\phi_s)}$ modulation of $A_{UT}(\rho^0)$ for HERMES (left) and COMPASS 
(right) kinematics. The results of our calculations, shown as solid lines, are compared to 
the data from \ci{hermes-aut} and \ci{compass-aut-sinphimphis}.}
\label{fig:sinphimphis}
\begin{center}
\includegraphics[width=0.37\tw]{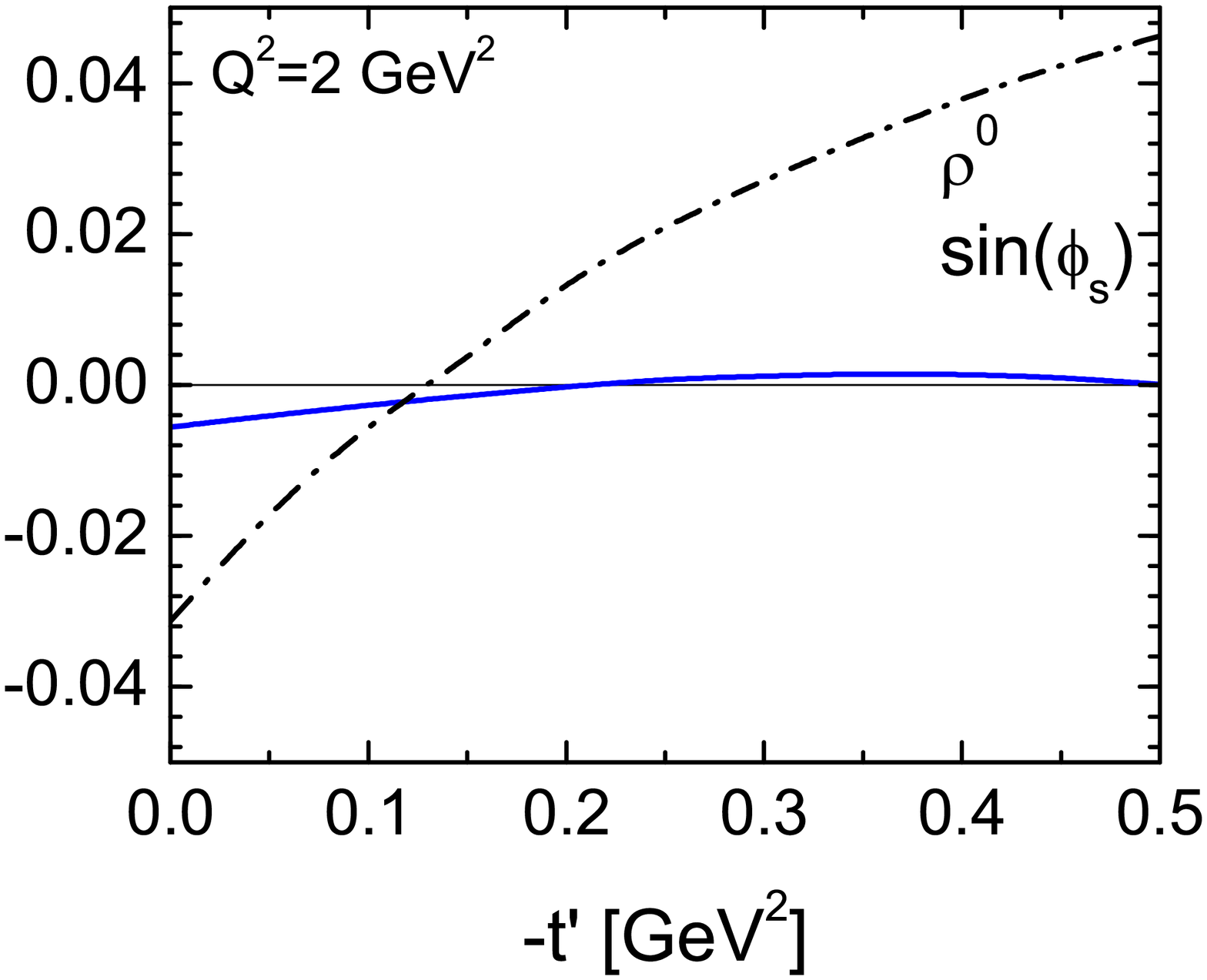}\hspace*{0.05\tw}
\includegraphics[width=0.37\tw]{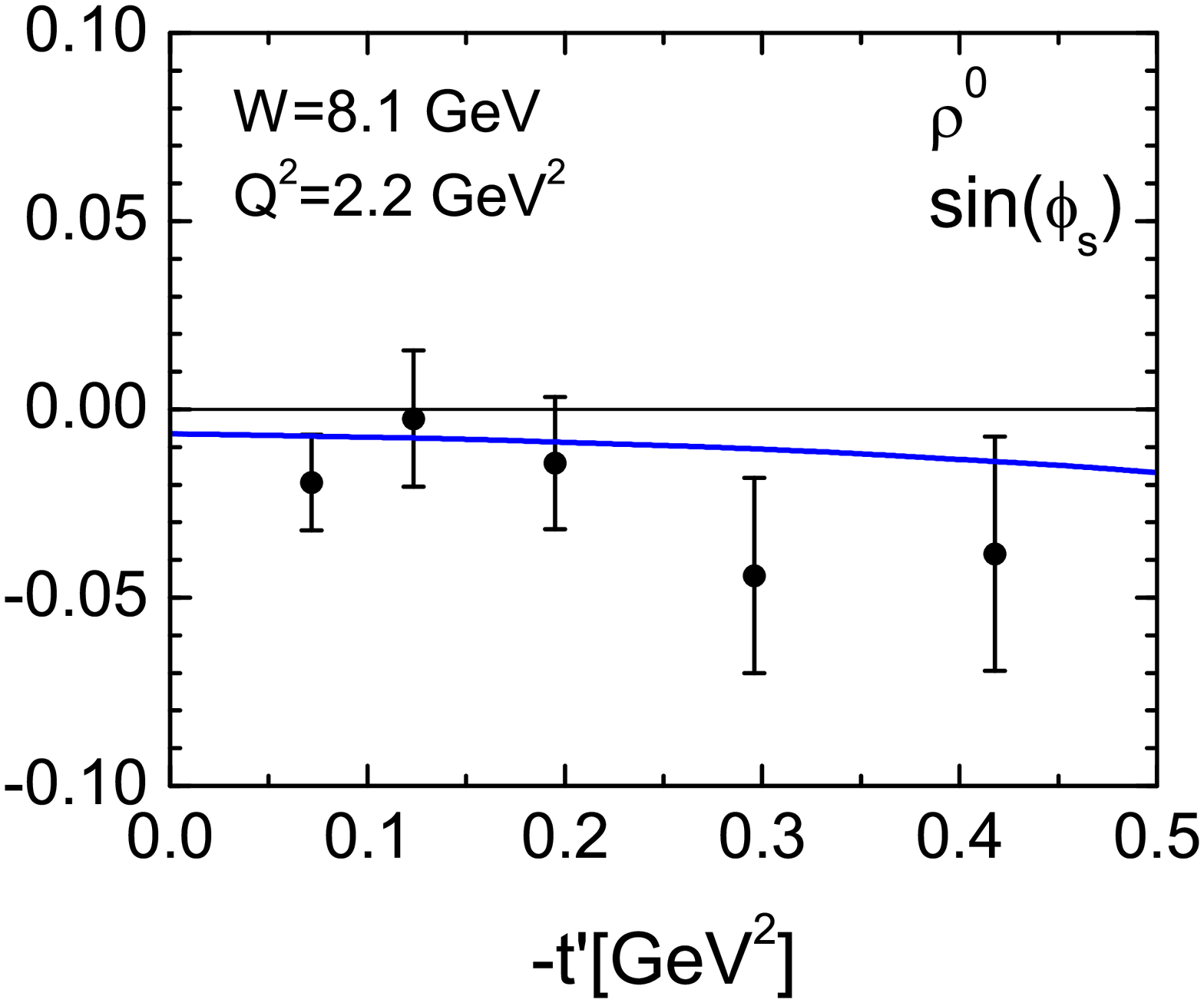}
\end{center}

\vspace*{-0.05\tw}
\caption{As Fig.\ \ref{fig:sinphimphis} but for the $\sin(\phi_s)$ modulation. For COMPASS
kinematics (right) the factor $\sqrt{\varepsilon (1+\varepsilon)}$ is taken out. Data are
taken from \ci{compass-aut}. Left: Predictions at $W=5\,\gev$ (solid line) and $3\,\gev$ 
(dash-dotted line).}
\label{fig:sinphis}
\end{figure}

The $\sin(\phi-\phi_s)$ modulation of $A_{UT}$ has been measured by the HERMES
\ci{hermes-aut} and COMPASS collaborations \ci{compass-aut-sinphimphis} for
$\rho^0$ leptoproduction. In \ci{GK4} this asymmetry has already been investigated by us 
and shown to be in reasonable agreement with experiment. However, the transversity GPDs 
were not taken into account in this analysis. The present analysis reveals that their 
contribution to the $\sin(\phi-\phi_s)$ modulation is small, the 
$\langle E\rangle^*_{LL}\langle H\rangle_{LL}^{\phantom{*}}$ and 
$\langle E\rangle^*_{TT}\langle H\rangle_{TT}^{\phantom{*}}$ interference terms are
dominant~\footnote
{Note that the $\sin(\phi-\phi_s)$ modulation is the only one that has a pure
leading-twist contribution, namely $\langle E\rangle^*_{LL}\langle H\rangle_{LL}^{\phantom{*}}$.}. 
This is obvious from the $\sin{(\phi+\phi_s)}$ modulation shown in Fig.\ \ref{fig:aut-compass} 
which is related to just the same interference term, 
$\langle \bar{E}_T\rangle_{LT}^*\langle H_T\rangle_{LT}^{\phantom{*}}$, as the contributions 
from the $\gamma^*_T\to V^{\phantom{*}}_L$ transitions to the $\sin(\phi-\phi_s)$ modulation. 
A small, almost zero $\sin(\phi+\phi_s)$ modulation is in agreement with experiment 
\ci{compass-aut} within errors. Hence, the results presented 
in \ci{GK4} essentially remain valid. For completeness we show these results here again, see 
Fig.\ \ref{fig:sinphimphis}. The $\sin(2\phi- \phi_s)$ modulation which is also shown in 
Fig.\ \ref{fig:aut-compass}, is very small in agreement with experiment \ci{compass-aut}.
It is related to the $\langle \bar{E}_T\rangle^*_{LT} \langle E\rangle_{LL}^{\phantom{*}}$
interference term. The $\sin(3\phi-\phi_s)$ modulation is strictly zero in our approach 
since it is related to interference terms  with the neglected  ${\cal M}^V_{0-,-+}$ 
and $\gamma^*_T\to V^{\phantom{*}}_{-T}$ amplitudes. At large values of $-t^\prime$ this is not
in good agreement with the COMPASS data \ci{compass-aut}, the deviations
amount to a bit more than one standard deviation.

\newpage
\clearpage

The $\sin(\phi_s)$ modulation is related to interference terms between transversity GPDs and 
$H, E$ like the SDME $r^5_{00}$ but with interchanged $H$ and $E$ contributions
\be
A_{UT}^{\sin(\phi_s)} \sim {\rm Im} \Big[\langle \bar{E}_T\rangle^*_{LT} 
                                             \langle E\rangle_{LL}^{\phantom{*}} 
                      -\langle H_T\rangle^*_{LT} \langle H\rangle_{LL}^{\phantom{*}}\Big]\,.  
\label{eq:sinphis}
\ee
The first term makes up the $\sin(2\phi-\phi_s)$ modulation and we already
know that it is very small, see Fig.\ \ref{fig:aut-compass}. The second term
in \req{eq:sinphis} is larger since, as we already mentioned, $\langle H\rangle_{LL}$ 
is much larger than $\langle E\rangle_{LL}$. This term is an interference term of two 
helicity non-flip amplitudes and is therefore not forced to vanish for forward 
scattering by angular momentum conservation in contrast to the first term which behaves 
$\propto t'$ for $t'\to 0$. Results for the $\sin(\phi_s)$ modulation are shown in 
Fig.\ \ref{fig:sinphis}. For COMPASS kinematics it is negative and amounts to about 
0.02 in absolute value. This is in reasonable agreement with experiment given that our 
results are only estimates and do not represent detailed fits to data. For HERMES 
kinematics the $\sin(\phi_s)$ modulation is very small while, at $W=3\,\gev$, we find for it
larger values and a zero at $t'\simeq -0.12\gev^2$.

\begin{figure}[t]
\begin{center}
\includegraphics[width=0.45\tw]{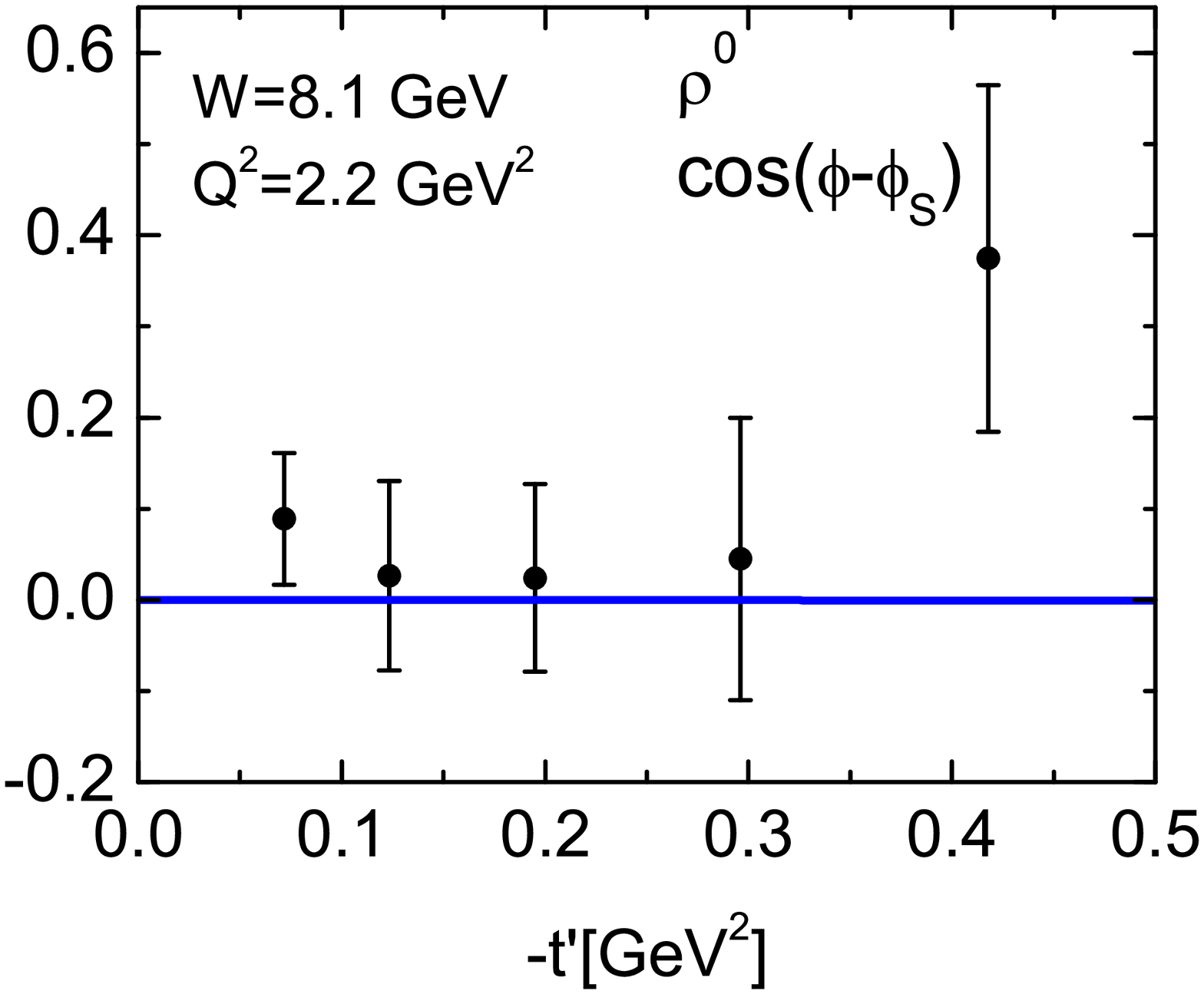}\hspace*{0.05\tw}
\includegraphics[width=0.45\tw]{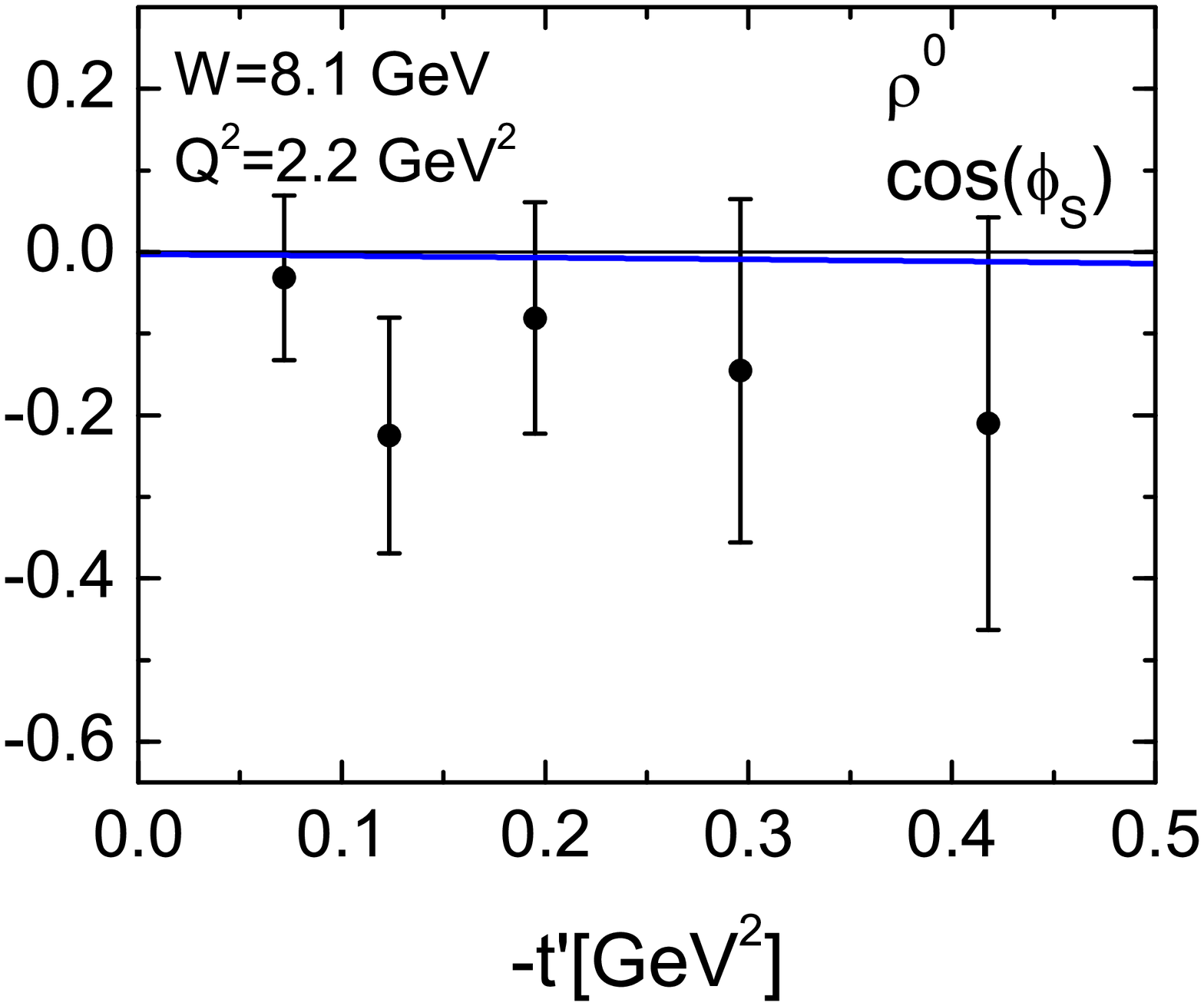}
\end{center}

\vspace*{-0.05\tw}
\caption{The $\cos{(\phi-\phi_s)}$ (left) and $\cos(\phi_s)$ (right) modulation of the
asymmetry $A_{LT}$ for $\rho^0$ leptoproduction. The prefactors $\sqrt{1-\varepsilon^2}$ 
and $\sqrt{\varepsilon(1+\varepsilon)}$ in \req{eq:alt} are taken out. The handbag results
are displayed as solid lines. Data are taken from \ci{compass-aut}.}
\label{fig:alt-compass}
\begin{center}
\includegraphics[width=0.45\tw]{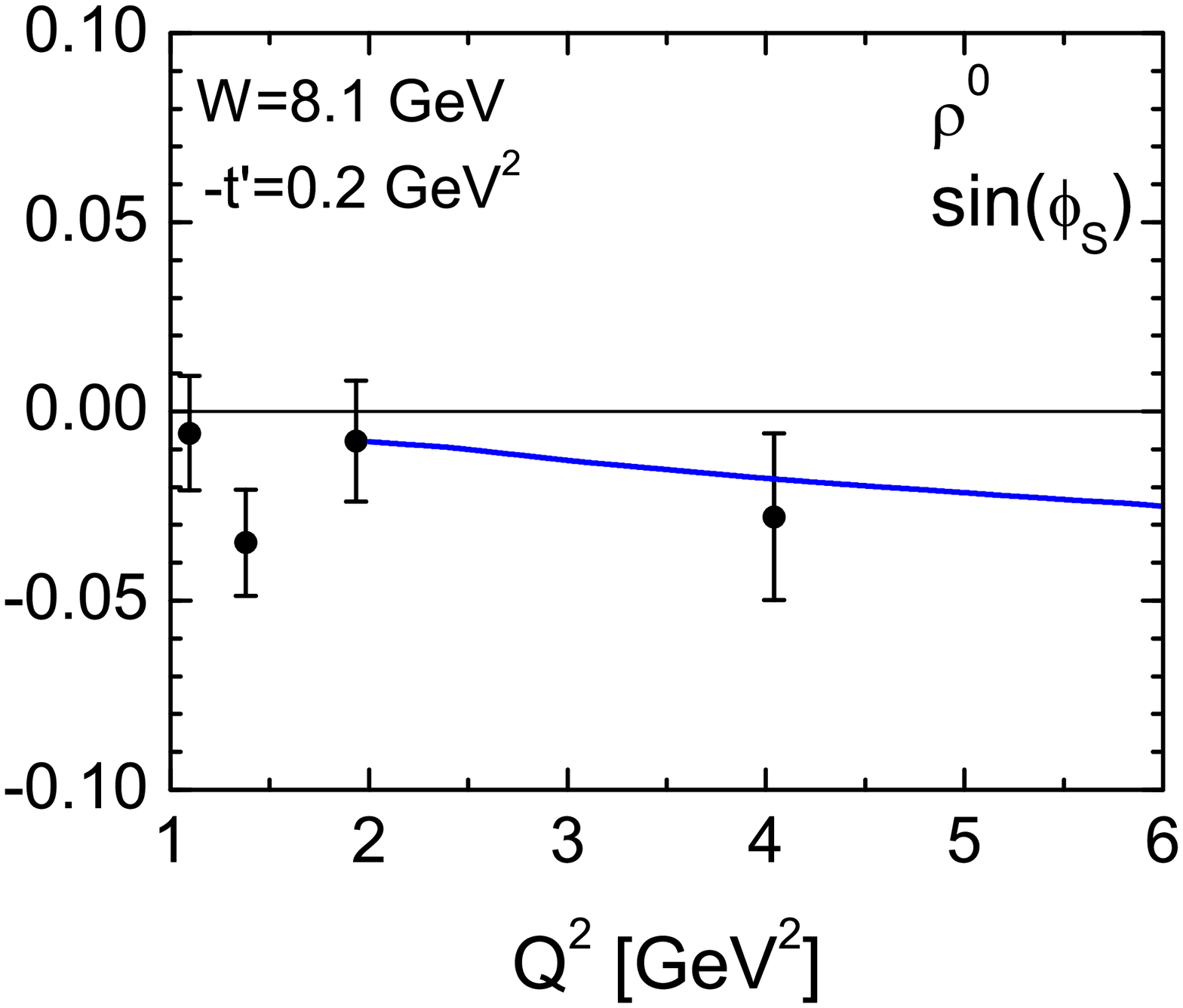}\hspace*{0.05\tw}
\includegraphics[width=0.44\tw]{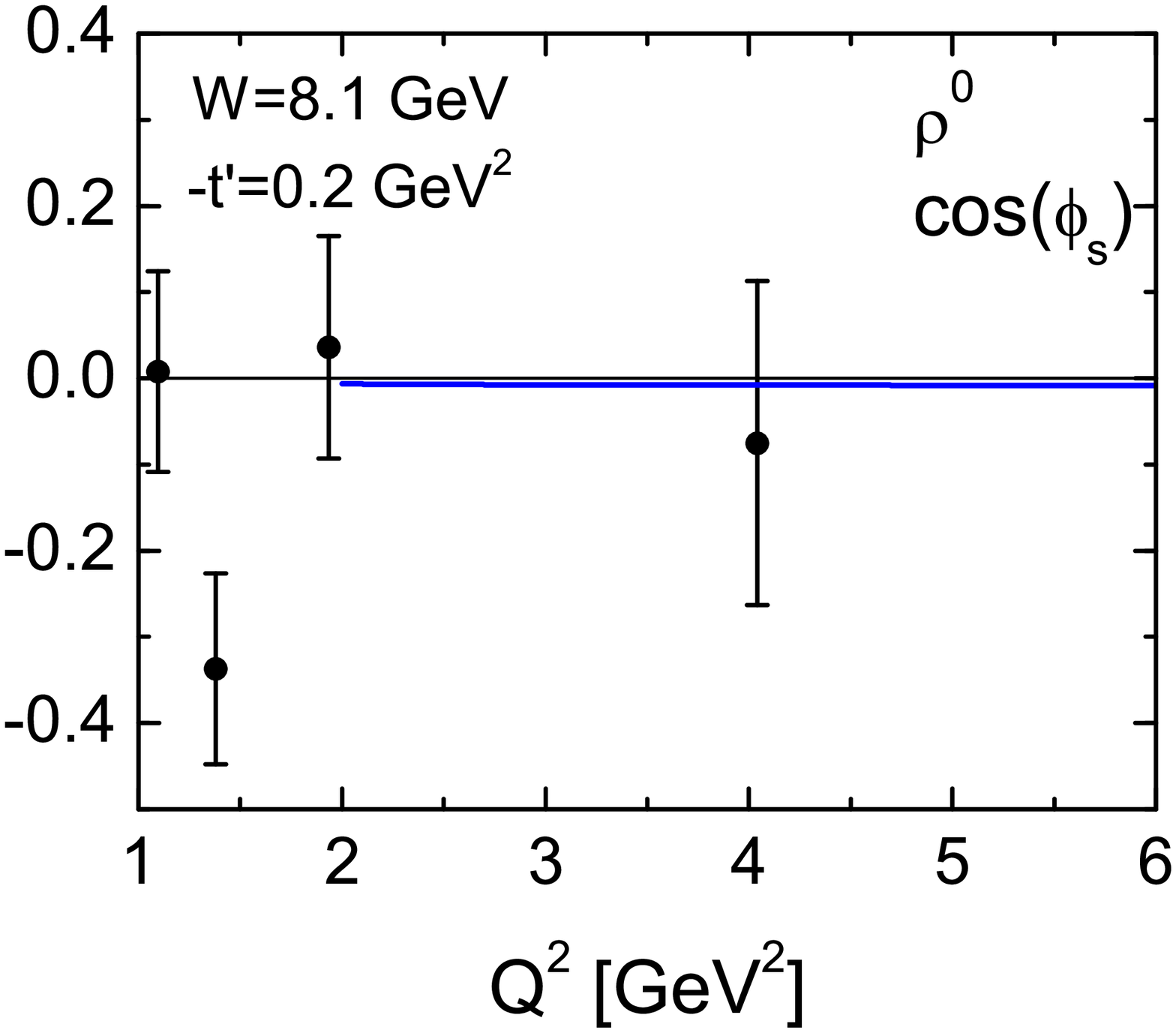}
\end{center}

\vspace*{-0.05\tw}
\caption{The $\sin(\phi_s)$ (left) and $\cos(\phi_s)$ (right) modulations for 
$\rho^0$ leptoproduction versus $Q^2$  at COMPASS kinematics.
The prefactors $\sqrt{\varepsilon (1\pm\varepsilon)}$ are taken out. The handbag results 
are shown as solid lines. Data are taken from \ci{compass-aut}.}
\label{fig:aut-alt-q2-compass}
\end{figure}

For a transversely polarized target and a longitudinally polarized beam various modulations
of the asymmetry $A_{LT}$ can be measured. In terms of helicity amplitudes the non-zero 
modulations read 
\ba
A_{LT}^{\cos{\phi_s}}(V)\sigma_0^V &=& \sqrt{\varepsilon(1-\varepsilon)}\,
   {\rm Re}\Big[{\cal M}^{V*}_{0+,++}{\cal M}_{0-,0+}^V - {\cal M}^{V*}_{0-,++}{\cal M}_{0+,0+}^V\Big]\,, 
                        \nn\\[0.2em]
A_{LT}^{\cos{(\phi -\phi_s)}}(V)\sigma_0^V &=& \sqrt{1-\varepsilon^2}\, 
                {\rm Re}\Big[{\cal M}^{V*}_{0-,++} {\cal M}_{0+,++}^V
                     -2 {\cal M}^{VN*}_{+-.++}{\cal M}^{VU}_{++,++}\Big] \nn\\[0.2em]
A_{LT}^{\cos{(2\phi-\phi_s)}}(V)\sigma_0^V &=& -\sqrt{\varepsilon(1-\varepsilon)}\,
                  {\rm Re}\Big[{\cal M}^{V*}_{0+,++} M_{0-,0+}^V\Big]\,.  
\label{eq:alt}
\ea
Leaving aside the $\varepsilon$-dependent prefactors in \req{eq:alt} the modulations 
$\cos(\phi_s)$ and $\cos{(2\phi-\phi_s)}$ of $A_{LT}$ are related to the same combinations 
of helicity amplitudes as the corresponding modulations of $A_{UT}$ except that the 
imaginary parts are to be substituted by the real parts. The  $\cos(\phi-\phi_s)$ modulation 
contains the real part of the $\langle\bar{E}_T\rangle_{LT}^*\langle H_T\rangle_{LT}^{\phantom{*}}$ 
interference term as in $A_{UT}^{\sin{(\phi-\phi_s)}}$ and a 
$\langle E\rangle^*_{TT}\langle \widetilde{H}\rangle_{TT}$ term. The imaginary part of the 
$\langle\bar{E}_T\rangle_{LT}^*\langle H_T\rangle_{LT}^{\phantom{*}}$ 
interference term also controls the $\sin(\phi+\phi_s)$ modulation of $A_{UT}$. In our handbag 
approach the $\cos(\phi-\phi_s)$ and $\cos{(2\phi-\phi_s)}$ 
modulations are very small as are the $\sin(\phi+\phi_s)$ and $\sin(2\phi-\phi_s)$ ones. 
The $\cos(\phi_s)$ modulation is similar in sign and size to $A_{UT}^{\sin(\phi_s)}$.
These results are in agreement with the COMPASS data \ci{compass-aut} within,
however, huge experimental errors. Two examples of the $A_{LT}$  modulations are shown 
in Fig.\ \ref{fig:alt-compass}. In contrast to the SDMEs discussed in Sect.\ 
\ref{sec:sdmes}, for which the contributions from $\bar{E}_T$ are dominant, the only 
substantial contribution from the transversity GPDs to the asymmetries $A_{UT}$ and 
$A_{LT}$ is that from $H_T$. The $\cos(\phi_s)$ modulation is rather strongly influenced 
by $H_T^s$. Without it this modulation would be positive in conflict with experiment.

The COMPASS collaboration \ci{compass-aut} has also measured the $Q^2$ and the $\xbj$ 
dependence of the asymmetries $A_{UT}$ and $A_{LT}$ for $\rho^0$ leptoproduction. In Fig.\
\ref{fig:aut-alt-q2-compass} we confront the $Q^2$ dependence of these data with
our results. Again agreement is to be seen within experimental errors. 
Results of similar quality are obtained for the $\xbj$ dependence. 
The calculated asymmetries are often very small and hardly to distinguish from zero 
in the plots.

\subsection{Predictions for other vector mesons}
Estimates of the unseparated cross sections for $\omega$, $\rho^+$ and $K^{*0}$ 
leptoproduction without the $\gamma^*_T\to V_L^{\phantom{*}}$ transitions have been 
given in \ci{GK4}. For the case of the $\omega$ the new contributions increase 
the cross section a little, about $2-3\%$ as is the case for the $\rho^0$ channel. 
On the other hand, for $\rho^+$ and $K^{*0}$ production the cross sections 
increase by about $20-30\%$ as compared to the estimates presented in \ci{GK4} 
(the quoted values are for COMPASS kinematics). Worth to mention is that the 
$\omega$ cross section is about an order of magnitude smaller than the $\rho^0$ one. 
Due to the absence of the contributions from $H$ for gluons the $\rho^+$ and $K^{*0}$
cross sections are even suppressed by about a factor of 100.

Since the $u$ and $d$ valence quark GPDs of $\bar{E}_T$ have the same sign and roughly 
the same strength (see Tab.\ \ref{tab:1})  a partial cancellation of both the 
contributions occur for $\omega$ and $\rho^+$ production as a consequence of
the flavor composition of these mesons, see \req{eq:flavor} and \req{eq:non-diagonal}. 
The resulting rather small contribution from $\bar{E}_T$ is however compensated to some 
extent by smaller cross sections. These properties result in substantially different SDMEs. 
As examples we show $r_{00}^1$ and $r_{00}^5$ in Fig.\ \ref{fig:r100-compass} for typical 
COMPASS kinematics. As is to be seen from this figure both the SDMEs, $r_{00}^1$ (in 
absolute value) and $r_{00}^5$, are slightly larger for the $\omega$ channel than for the 
$\rho^0$ one. For the case of the $\rho^+$ the SDMEs are noticeably larger. Even 
strikingly larger SDMEs are found for the $K^{*0}$ channel. This is so because only 
$\bar{E}_T^{d_v}$ contributes and the cross section is very small. We note in passing
that the HERMES collaboration \ci{marianski} has shown preliminary data on the SDME 
for $\omega$ production at the DIS 2013 ($W=5\,\gev$, $Q^2=2\,\gev^2$). For the SDMEs 
under control of the 

\newpage
\clearpage

\noindent transversity GPDs we find fair agreement between these data and
the results from our handbag approach. 

Since $H_T$ for $u$ and $d$ valence quarks have opposite signs (see Tab.\ \ref{tab:1})  
a partial cancellation of the two contributions takes place for the $\rho^0$
channel while they add for $\omega$ and $\rho^+$ production. Moreover, the
absence of the contribution from $H$ for gluon leads to very different
relative phases between $\langle H_T\rangle_{LT}$ and $\langle H\rangle_{LL}$ for $\rho^+$ 
and $K^{*0}$ production. Thus, larger modulations of $A_{UT}$ and $A_{LT}$ are 
to be expected in particular for the $\rho^+$ and $K^{*0}$ channels than for 
$\rho^0$ production. Indeed for the $\sin(\phi_s)$ and $\cos(\phi_s)$ modulations 
displayed in Fig.\ \ref{fig:aut-alt-compass}, this pattern is clearly seen.

Predictions for $A_{UT}^{\sin(\phi-\phi_s)}$  for $\omega, K^{*0}$ and $\rho^+$
leptoproduction are already given in \ci{GK4}. With regard to the fact that the 
contributions from the $\gamma^*_T\to V^{\phantom{*}}_L$ amplitudes play only a 
minor role for this modulation, the results presented in \ci{GK4} remain unchanged 
practically. The $\sin(\phi-\phi_s)$ modulation is much larger for $\omega$, $\rho^+$ 
and $K^{*0}$ channels than for $\rho^0$ production. The largest asymmetry  
$A_{UT}^{\sin(\phi-\phi_s)}$ is found for $\rho^+$ production. It also exhibites a very 
different $t'$-dependence and opposite sign than for the other vector meson channels. 
This is a consequence of the large helicity flip amplitude ${\cal M}_{0-,0+}$ which is 
related to the GPD $E$. The amplitude ${\cal M}_{0+,0+}$ is not much larger than the 
flip amplitude for this channel since the gluon GPD does not contribute and because 
of the cancellation in the flavor combination of $u$ and $d$ valence quarks for $H$  
while, for $E$, both the contributions add. For further details of this asymmetry it 
is referred to \ci{GK4}. For $\phi$ leptoproduction all modulations of $A_{UT}$ and 
$A_{LT}$ as well as the SDMEs given in \req{eq:sdme} are very small since the strange 
transversity GPDs $H_T$ and $\bar{E}_T$ are small. On the other hand, experimental data 
on these observables may allow for a better determination of these GPDs.

\subsection{Longitudinal polarization}

More asymmetries can be measured with a longitudinally polarized beam and/or 
target. Though there is no data on such asymmetries available as yet except of a 
few data points for exclusive $\rho^0$ production on the proton \ci{SMC,hermes03}
and the deuteron \ci{compass07} with however very large errors, we will
discuss them briefly here. Using the simplifications discussed is Sect.\ 
\ref{sec:handbag} (see Eq.\ \req{eq:TL-simple}) and ignoring again the difference
between the directions of the virtual photon and the incoming lepton, we find
the following non-zero observables

\newpage
\begin{figure}[t]
\begin{center}
\includegraphics[width=0.45\tw]{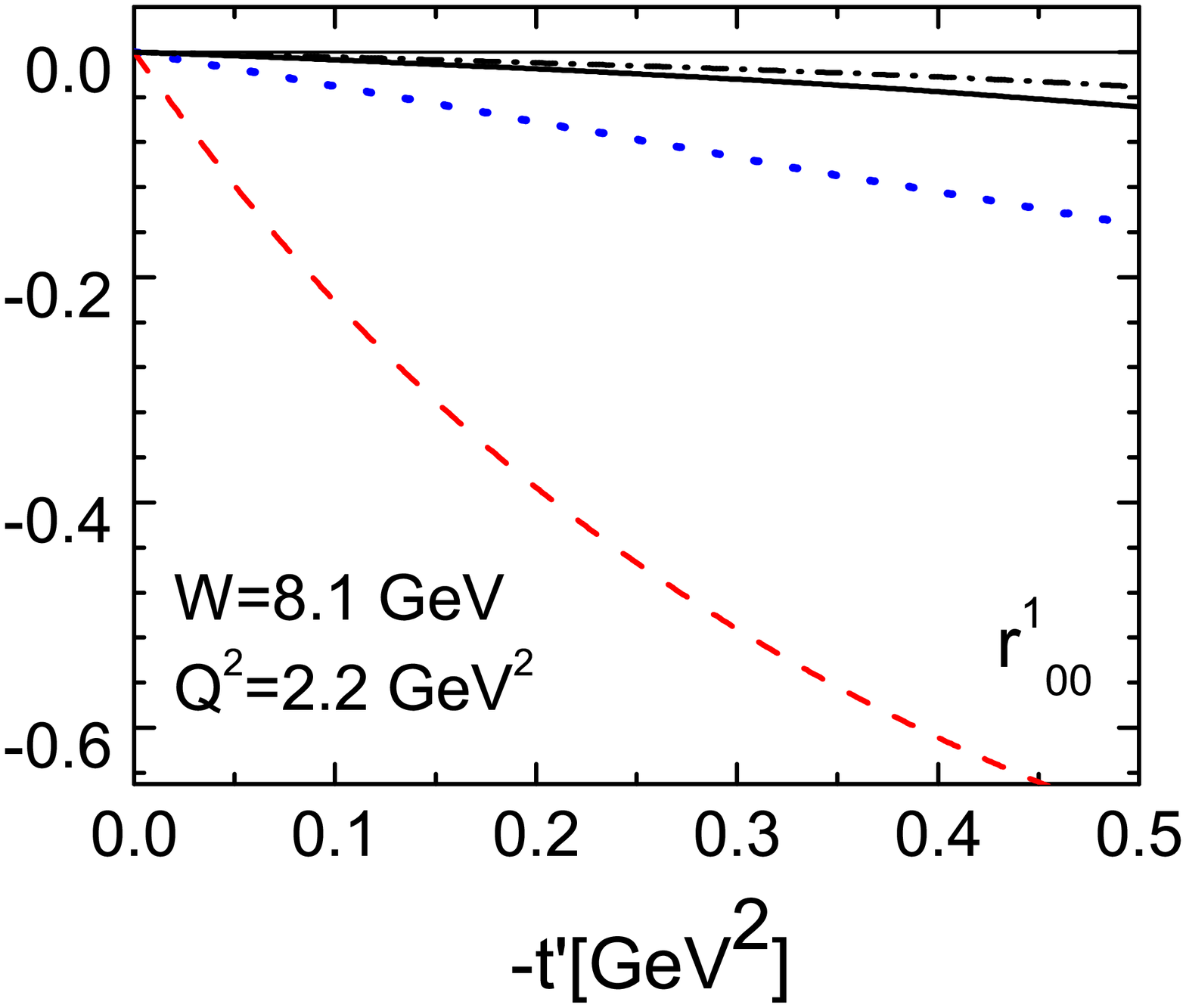}
\includegraphics[width=0.45\tw]{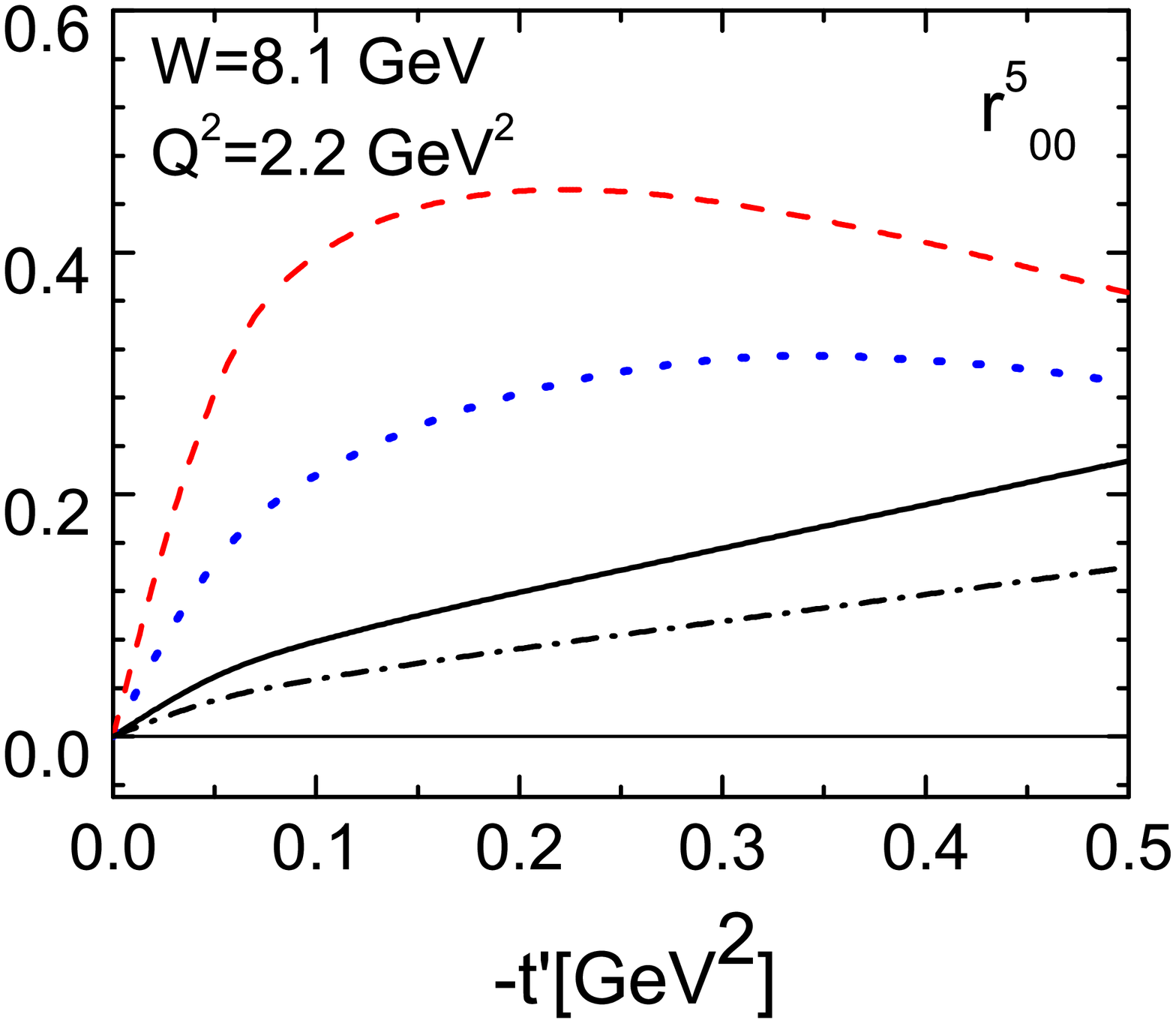}
\end{center}

\vspace*{-0.05\tw}
\caption{Predictions for the SDME $r_{00}^1$ (left) and $r_{00}^5$ (right) for 
$\omega$ (solid), $\rho^0$ (dash-dotted), $\rho^+$ (dotted) 
and $K^{*0}$ (dashed line) leptoproduction at COMPASS kinematics.} 
\label{fig:r100-compass}
\begin{center}
\includegraphics[width=0.45\tw]{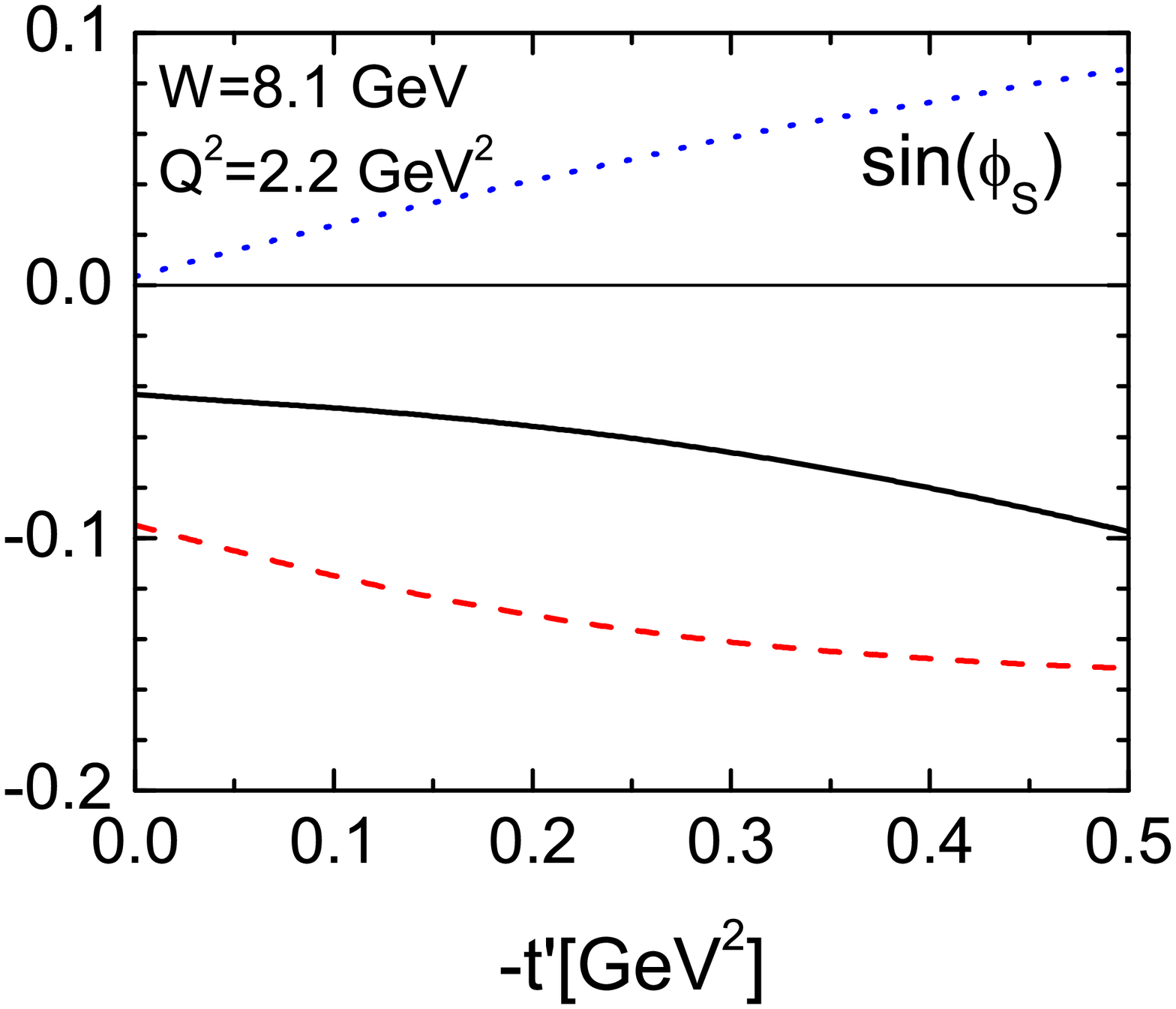}
\includegraphics[width=0.45\tw]{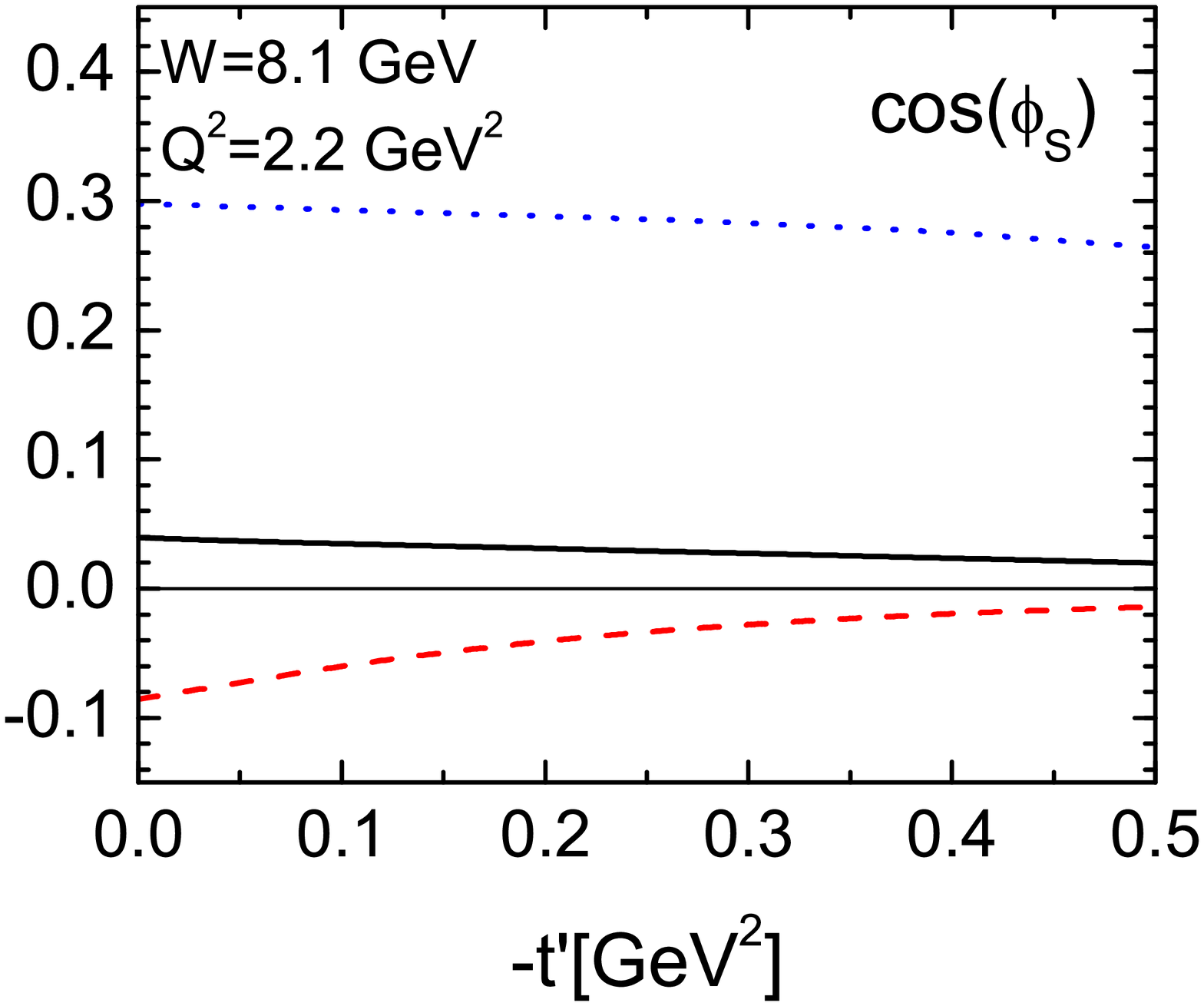}
\end{center}

\vspace*{-0.05\tw}
\caption{Predictions for $A_{UT}^{\sin(\phi_s)}$ (left) and $A_{LT}^{\cos(\phi_s)}$ (right) 
for $\omega$ (solid line), $\rho^+$ (dotted line) and $K^{*0}$ (dashed line) leptoproduction
at a typical COMPASS kinematics. The prefactors $\sqrt{\varepsilon (1\pm\varepsilon)}$ 
are taken out. The handbag results are shown as solid lines. Data are taken from \ci{compass-aut}.}
\label{fig:aut-alt-compass}
\end{figure}
\clearpage

\ba
A_{LU}^{\sin(\phi)}(V)\sigma_o^V&=&-\sqrt{\varepsilon(1-\varepsilon)}\, {\rm Im}
                 \Big[2{\cal M}_{0+,++}^{V*}{\cal M}_{0+,0+}^V 
                      + {\cal M}_{0-,++}^{V*}{\cal M}_{0-,0+}^V\Big]\,, \nn\\
A_{UL}^{\sin(\phi)}(V)\sigma_o^V&=&-\sqrt{\varepsilon(1+\varepsilon)}\, {\rm Im}
                 \Big[{\cal M}_{0-,++}^{V*}{\cal M}_{0-,0+}^V\Big]\,, \nn\\ 
A_{LL}^{\cos(0\phi)}(V)\sigma_0^V&=& \phantom{-}\sqrt{1-\varepsilon^2}\left\{
                    2{\rm Re}\Big[{\cal M}_{++,++}^{VN*}{\cal M}_{++,++}^{VU}\Big]
                    + \frac12 |{\cal M}_{0-,++}^V|^2 \right\}\,,\nn\\
A_{LL}^{\cos(\phi)}(V)\sigma_0^V&=&-\sqrt{\varepsilon (1-\varepsilon)}\,{\rm Re}
                        \Big[{\cal M}_{0-,++}^{V*}{\cal M}_{0-,0+}^V\Big]\,.
\ea
The asymmetry $A_{LU}$ measures the imaginary part of the same interference 
term as the SDME $r_{00}^5$. Thus, we expect an $A_{LU}$, divided by 
$\sqrt{2\varepsilon(1-\varepsilon)}$, slightly smaller than $r_{00}^5$.
As we discussed in Sect.\ \ref{sec:sdmes} the term ${\cal M}^*_{0-,++}{\cal M}_{0-,0+}$ 
being related to the GPDs $H_T$ and $\tilde E$, is very small with the consequence 
of small $A_{UL}$ and $A_{LL}^{\cos(\phi)}$ at least for $\rho^0$ and $\omega$
production. The asymmetry $A_{LL}^{\cos(0\phi)}$ receives a contribution
from the $\gamma^*_T\to V_T^{\phantom{*}}$ amplitudes, i.e.\ from the interference term of 
$\langle H\rangle_{TT}$ and $\langle \widetilde{H}\rangle_{TT}$. There is also a 
contribution to it from the transversity GPD $H_T$ which was not taken into account in 
our previous work \ci{GK3,GK1} where we already analysed $A_{LL}$ for $\rho^0$ 
production. Since in our approach $|{\cal M}_{0-,++}|<|{\cal M}_{0+,++}|$ the
additional term is smaller than $-r^1_{00}/2$. With regard to our results on the
SDME $r_{00}^1$ displayed in Figs.\ \ref{fig:sdme1} and \ref{fig:r100-compass},
and those on the interference of the $\gamma_T^*\to V_T^{\phantom{*}}$ amplitudes
presented in \ci{GK3} we find a small asymmetry  $A_{LL}^{\cos(0\phi)}$
for $\rho^0$ and $\omega$ production at COMPASS kinematics. However, a revision
of the parametrization of $\widetilde{H}$ given in \ci{GK3} seems to be advisable.

\section{Summary}

The role of transversity GPDs in vector-meson leptoproduction is investigated. 
It is argued that these GPDs control the $\gamma^*_T\to V_L^{\phantom{*}}$ transition 
amplitudes and constitute a twist-3 effect consisting of leading-twist GPDs in 
combination with twist-3 meson \wf s. As compared to the asymptotically 
leading $\gamma^*_L\to V_L^{\phantom{*}}$ amplitudes the 
$\gamma^*_T\to V_L^{\phantom{*}}$ ones are suppressed by $m_V/Q$. In contrast to pion 
leptoproduction the $\gamma^*_T\to V_L^{\phantom{*}}$ amplitudes do not affect the 
unpolarized cross sections considerably; they only influence markedly some of the 
SDMEs and asymmetries measured with a transversely polarized target. In most cases 
they contribute via interferences with amplitudes under control of the helicity 
non-flip GPDs. For the estimates made in this work the parametrizations of the GPDs 
are taken from our previous work \ci{GK6,GK3}. The only new pieces introduced here 
are the sea-quark transversity GPDs. From this set of GPDs we evaluate various SDMEs 
and modulations of the asymmetries $A_{UT}$ and $A_{LT}$ and compare the results to 
HERMES \ci{hermes,hermes-aut} and COMPASS data \ci{compass-aut,compass-aut-sinphimphis}.
In general fair agreement with experiment is obtained. 

We stress that we do not attempt detailed fits of the transversity GPDs to the 
data on SDMEs and asymmetries. 
A precise calculation, including an error assessment, of the transversity effects
in leptoproduction of vector mesons is beyond feasibility at present. There are 
many uncertainties like the parameterization of the transversity GPDs or the 
exact treatment of the twist-3 contribution (e.g.\ the neglect of possible 
three-particle configurations of the meson state). Also higher-order perturbative 
corrections other than those included in the Sudakov factor and, implicitly, 
in the experimental electromagnetic form factor of the pion appearing in the 
pion-pole contribution to $\pi^+$ leptoproduction, are ignored. According to 
\ci{diehl-kugler} the NLO corrections to the leading-twist contribution are rather 
large for the cross sections for $Q^2\lsim 10\,\gev^2$. Further uncertainties occur 
for $K^{*0}$ production. In contrast to the case of the $\rho^+$ 
where the $p\to n$ transition GPDs are related to the diagonal proton ones by 
isospin symmetry, the proton - $\Sigma^+$ transition GPDs are connected to the 
proton GPDs by SU(3) flavor symmetry which is less accurate than isospin 
symmetry. The assumption of a flavor symmetric sea for all GPDs is also 
stronger for $K^*$ than for $\rho$ mesons. With regard to all these uncertainties 
we consider our investigation of leptoproduction of vector mesons 
as an estimate of the pertinent observables. The trends and magnitudes of 
the SDME and asymmetries are likely correct but probably not the details. 
Despite these uncertainties our estimates of transversity effects in
$\rho^0$ production for which data is available, work surprisingly well.
Data on other vector-meson channels are highly welcome; they will provide further 
checks of the transversity effects we are advocating. Such data may be provided
by COMPASS and by the upgraded Jlab in future. We are aware that such measurements
are a challenge for experimenters. We have shown only a few examples of SDMEs
and asymmetries for $\omega$, $\rho^+$ and $K^{*0}$ leptoproduction but we have
results for all observables discussed in this paper. Tables of these results can
be obtained from the authors on request.

{\bf Acknowledgements} 
We are grateful to Wolf-Dieter Nowak for drawing our attention to the problem
of interpreting the transverse target spin asymmetries and for his continous
interest in the ongoing analysis. This work is supported  in part by the Russian 
Foundation for Basic Research, Grant 12-02-00613 and by the Heisenberg-Landau
program and by the BMBF, contract number 05P12WRFTE.\\


\end{document}